\documentclass[12pt]{article}
\usepackage{geometry,epstopdf}
\usepackage{graphicx}

\usepackage{amsfonts}

\usepackage{cite}
\def \d  {{\rm d}}
\def \im {{\rm i}}
\def \ex {{\rm e}}

\def \boldk {\mbox{\boldmath$k$}}
\def \boldl {\mbox{\boldmath$l$}}
\def \boldm {\mbox{\boldmath$m$}}

\newcommand{\tgh}{\tanh}
\newcommand{\sign}{\mathop{\rm sign}\nolimits}
\newcommand{\om}{\Omega}
\def \rovno {\!\!&=&\!\!}

\begin{document}

\title{\bf Non-expanding Pleba\'nski--Demia\'nski space-times}

\author{
 J. Podolsk\'y$^1$\thanks{E--mail: {\tt podolsky(at)mbox.troja.mff.cuni.cz}},
 \
 O. Hru\v{s}ka$^1$\thanks{E--mail: {\tt HruskaOndrej(at)seznam.cz}}
 \ and
 J. B. Griffiths$^2$\thanks{E--mail: {\tt j.b.griffiths(at)icloud.com}}
\\ \\
\small $^1$ Institute of Theoretical Physics, Charles University, Prague\\
\small V Hole\v{s}ovi\v{c}k\'ach 2, 180\,00 Prague 8, Czech Republic.\\
\small $^2$ Retired. }

\date{\today}
\maketitle

\begin{abstract}
\noindent The aim of this work is to describe the complete
family of non-expanding Pleba\'nski--Demia\'nski type~D
space-times and to present their possible interpretation. We
explicitly express the most general form of such
(electro)vacuum solutions with any cosmological constant, and
we investigate the geometrical and physical meaning of the
seven parameters they contain. We present various metric forms,
and by analyzing the corresponding coordinates in the
weak-field limit we elucidate the global structure of these
space-times, such as the character of possible singularities.
We also demonstrate that members of this family can be
understood as generalizations of classic $B$-metrics. In
particular, the $BI$-metric represents an external
gravitational field of a tachyonic (superluminal) source,
complementary to the $AI$-metric which is the well-known
Schwarzschild solution for exact gravitational field of a
static (standing) source.
\end{abstract}

\newpage

\section{Introduction}
\label{intro}

\begin{figure}[t!]
\centerline{\includegraphics[scale=1]{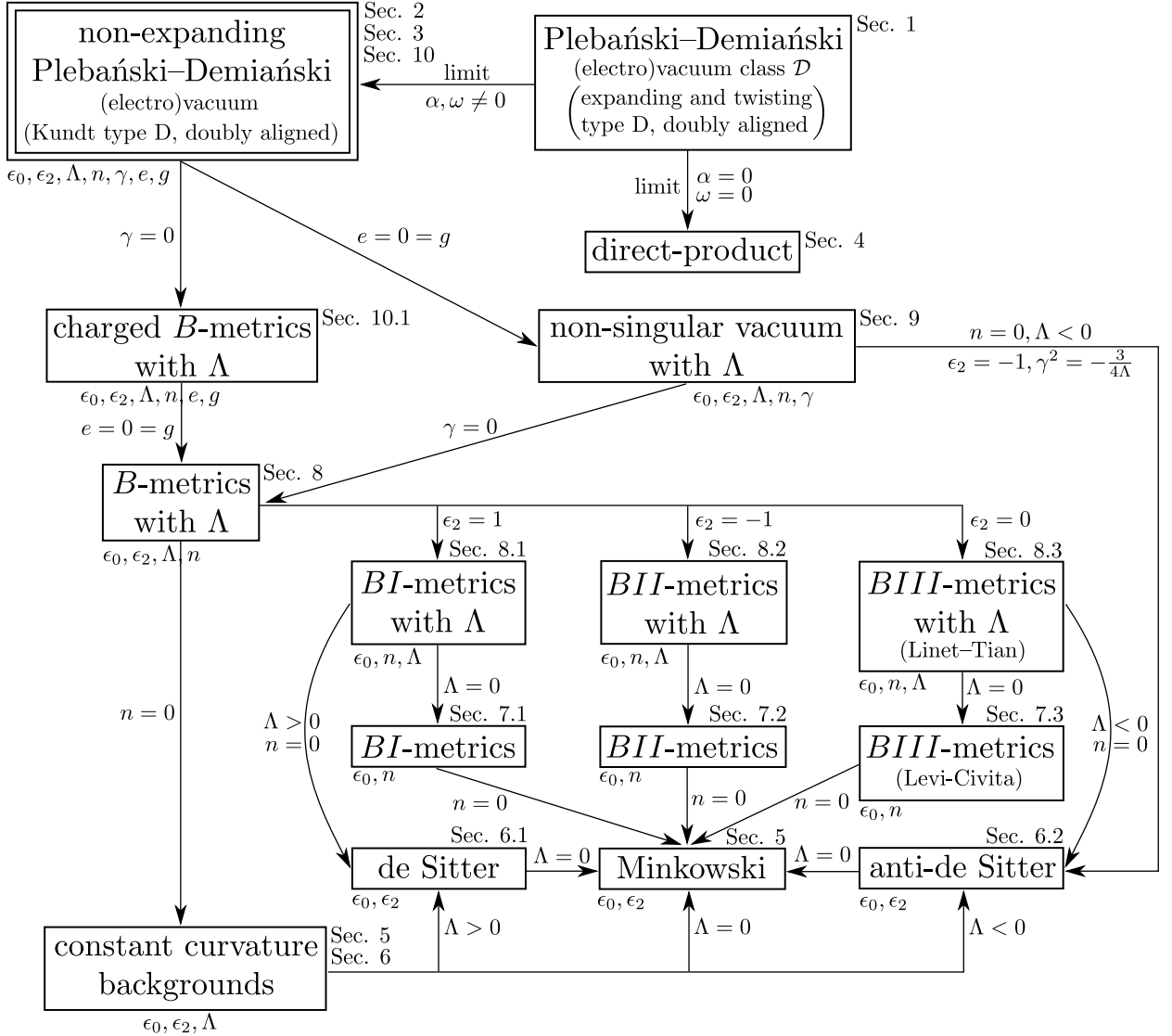}}
\caption{ \small Schematic diagram of the structure of the complete family of
non-expanding Pleba\'nski--Demia\'nski space-times. These are (electro)vacuum solutions
of the Einstein(--Maxwell) equations with any cosmological constant $\Lambda$
(and aligned electromagnetic field). All solutions belong to the Kundt class, and
their gravitational field is of algebraic type~D. By setting any of the seven
independent parameters $\epsilon_0$, $\epsilon_2$, $\Lambda$, $n$, $\gamma$, $e$, $g$
to zero, various specific subfamilies are obtained, such as the $B$-metrics and
background spaces of constant curvature (Minkowski, de~Sitter, anti-de~Sitter).
Each of these subfamilies is analyzed in a specific Section of this contribution,
as also indicated in the diagram.}
\label{diagram}
\end{figure}

The famous class of Pleba\'nski--Demia\'nski space-times is the
most general family of exact solutions of the
Einstein(--Maxwell) equations with any value of the
cosmological constant $\Lambda$, whose gravitational fields are
of algebraic type~D and electromagnetic fields are doubly
aligned. The class includes two distinct families according to
whether or not the repeated principal null directions are
expanding. In the \emph{expanding case} they involve nine
distinct parameters, and include a family of generalized black
hole space-times. In the \emph{non-expanding case}, however,
there are fewer parameters. We will show in
Sections~\ref{intro}--\ref{furthercases} that the complete
family of such solutions involves seven parameters, namely
$\epsilon_0$, $\epsilon_2$, $\Lambda$, $n$, $\gamma$, $e$, and
$g$. The geometrical and/or physical meaning of these
parameters will be clarified in
Sections~\ref{Minkowski}--\ref{interp_eg}. Moreover, by setting
any of these parameters to zero, specific subfamilies are
directly obtained, namely the $B$-metrics and their
generalizations to include the cosmological constant and an
aligned electromagnetic field. A diagram summarizing all these
subfamilies and their mutual relations is presented in
Fig.\ref{diagram}.

The complete class of the Pleba\'nski--Demia\'nski solutions
\cite{PleDem76} can be conveniently expressed in terms of the
line element \cite{GriPod06b,GriPod09}
  \begin{eqnarray}
  &&\hskip-3pc\d s^2=\frac{1}{(1-\alpha pr)^2} \Bigg[
 -\frac{{\cal Q}^\ex}{ r^2+\omega^2p^2}(\d\tau-\omega p^2\d\sigma)^2
 +\frac{r^2+\omega^2p^2}{{\cal Q}^\ex}\,\d r^2 \nonumber \\
  &&\hskip5pc
  +\frac{{\cal P}^\ex}{r^2+\omega^2p^2}(\omega\d\tau+r^2\d\sigma)^2
  +\frac{r^2+\omega^2p^2}{{\cal P}^\ex}\,\d p^2 \Bigg], \label{PleDemMetric}
  \end{eqnarray}
  where
 \begin{equation}
 \begin{array}{rl}
 {\cal P}^\ex(p) \!\!&=k +2n\omega^{-1}p -\epsilon p^2 +2\alpha mp^3-\Big(\alpha^2(\omega^2 k+e^2+g^2)+{\textstyle\frac{1}{3}}\omega^2\Lambda \Big)\,p^4, \\[8pt]
 {\cal Q}^\ex(r) \!\!&=(\omega^2k+e^2+g^2) -2mr +\epsilon r^2 -2\alpha n\omega^{-1}r^3-(\alpha^2k+{\textstyle\frac{1}{3}}\Lambda)\,r^4,
 \end{array}
 \label{PQeqns}
 \end{equation}
 and $m$, $n$, $e$, $g$, $\Lambda$, $\epsilon$, $k$, $\alpha$, $\omega$
are arbitrary real parameters. This metric represents type~D
solutions for which the repeated principal null directions are
{\em shear-free}, \emph{expanding} and \emph{twisting}. Indeed,
adopting the null tetrad
 \begin{eqnarray}
 \boldk \rovno {\displaystyle \frac{1-\alpha pr}{\sqrt{2(r^2+\omega^2p^2)}}\left[
 \frac{1}{\sqrt {{\cal Q}^\ex}}\,\big(r^2\partial_\tau-\omega\partial_\sigma\big)
-{\sqrt {{\cal Q}^\ex}}\,\partial_r \right]} , \nonumber\\
 \boldl \rovno {\displaystyle \frac{1-\alpha pr}{\sqrt{2(r^2+\omega^2p^2)}}\left[
 \frac{1}{{\sqrt {{\cal Q}^\ex}}}\,\big(r^2\partial_\tau-\omega\partial_\sigma\big)
+{\sqrt {{\cal Q}^\ex}}\,\partial_r \right]} ,  \label{GeneralTetrad}\\
 \boldm \rovno {\displaystyle \frac{1-\alpha pr}{\sqrt{2(r^2+\omega^2p^2)}}\left[
 -\frac{1}{\sqrt{{\cal P}^\ex}}\,\big(\omega p^2\partial_\tau+\partial_\sigma\big)
+\im\,\sqrt{{\cal P}^\ex}\,\partial_p \right]} , \nonumber
 \end{eqnarray}
the spin coefficients are ${\kappa = 0 = \nu}$, ${\sigma = 0 = \lambda}$,
 \begin{equation}
 \rho =\sqrt{\frac{{\cal Q}^\ex}{2(r^2+\omega^2p^2)}}\, \frac{1+\im\,\alpha\omega p^2}{r+\im\,\omega p}= \mu \,,
 \label{spincoeffts}
 \end{equation}
and ${\tau=\pi}$, ${\epsilon=\gamma}$, ${\alpha=\beta}$ are
also non-zero. The congruences generated by $\boldk$ and
$\boldl$ are thus geodesic and shear-free, but have non-zero
expansion, and their twist is proportional to the
parameter~$\omega$. Using the tetrad (\ref{GeneralTetrad}), the
only non-trivial Weyl tensor component is
 \begin{equation}
 \Psi_2=-(m+\im\,n)\left(\frac{1-\alpha pr}{r+\im\,\omega p}\right)^3
+(e^2+g^2)\left(\frac{1-\alpha pr}{r+\im\,\omega p}\right)^3 \frac{1+\alpha pr}{r-\im\,\omega p}\,,
 \label{Weyl1}
 \end{equation}
confirming that these space-times are of algebraic type~D with the repeated principal null directions $\boldk$ and $\boldl$. Apart from $\Lambda$, the only non-zero component of the Ricci tensor is
 \begin{equation}
 \Phi_{11}= \frac{1}{2}\,(e^2+g^2)\,\frac{(1-\alpha pr)^4}{(r^2+\omega^2p^2)^2}\,,
 \label{Ricci1}
 \end{equation}
where $e$ and $g$ are the electric and magnetic charges of the
source, respectively. Both principal null directions of the
non-null electromagnetic field are thus aligned with the
repeated principal null directions of the gravitational field.
Clearly, there is a curvature singularity at ${r=0=\omega p}$.
In general, this is surrounded by horizon(s) which are roots of
the function ${{\cal Q}^\ex(r)}$. In fact, the expanding metric
(\ref{PleDemMetric}), (\ref{PQeqns}) includes a \emph{large
family of black holes with various physical parameters}, such
as the mass $m$, Kerr-like rotation $a$, NUT parameter $l$
(related to the twist parameter $\omega$ and $n$), cosmological
constant $\Lambda$, electromagnetic charges ${e, g}$ and
acceleration $\alpha$, see \cite{GriPod06b,GriPod09} for more
details.

Interestingly, \emph{non-expanding Pleba\'nski--Demia\'nski}
type~D space-times can be obtained from the line element
(\ref{PleDemMetric}), which represents expanding space-times,
by applying specific ``degenerate'' transformation. Apart from
the exceptional case of direct-product geometries
\cite{GriPod06b,GriPod09}, see Section~\ref{furthercases}
below, the \emph{general family} of such solutions is obtained
by applying the transformation
 \begin{equation}
 r = \gamma+\kappa q\,, \qquad
 \sigma = k_1\,y + \omega\kappa^{-1}\,t\,, \qquad
 \tau = k_2\,y-\gamma^2\kappa^{-1}\,t\,,
 \label{trans1B}
 \end{equation}
where $\gamma$ and $\kappa$ are arbitrary parameters, and taking the limit in which ${\kappa\to0}$. In this limit the function ${{\cal Q}^\ex}$ is \emph{rescaled to zero} as ${\kappa^2{\cal Q}}$, and the resulting line element takes the form
 \begin{equation}
 \d s^2= \frac{1}{(1-\alpha\gamma\, p)^2} \left[
 \varrho^2\Big(-{\cal Q}\,\d t^2 +\frac{1}{{\cal Q}}\,\d q^2 \Big)
  +\frac{{\cal P}^\ex}{\varrho^2} \Big((k_1\gamma^2+k_2\,\omega)\,\d y+2\gamma\omega\, q\,\d t \Big)^2
  +\frac{\varrho^2}{{\cal P}^\ex}\,\d p^2 \right] ,
 \label{nonExpMetric}
 \end{equation}
 where
\begin{eqnarray}
 \varrho^2 \rovno  \omega^2p^2+\gamma^2\,, \nonumber\\[3pt]
 {\cal Q} \rovno  \epsilon_0-\epsilon_2 q^2\,, \label{coeffnonexp}\\[3pt]
 {\cal P}^\ex \rovno  k +2n\omega^{-1}p -\epsilon p^2 +2\alpha mp^3
-\left(\alpha^2(\omega^2k+e^2+g^2)+{\textstyle\frac{1}{3}}\omega^2\Lambda\right)p^4\,, \nonumber
\end{eqnarray}
with an additional free constant  $\epsilon_0$ resulting from the limiting procedure, and
\begin{equation}
 \epsilon= -\epsilon_2 +6\alpha\gamma n\omega^{-1} +2\gamma^2(3\alpha^2k+\Lambda) \,.
 \label{epsilon}
\end{equation}
 The parameters of these solutions must also satisfy two further constraints, namely
 \begin{eqnarray}
 3m + \gamma(\epsilon_2-2\epsilon) +3\alpha\gamma^2n\omega^{-1} \rovno  0 \,, \label{m}\\[4pt]
 \omega^2k+e^2+g^2 -\gamma m +{\textstyle{1\over6}}\gamma^2(\epsilon+\epsilon_2) \rovno  0 \label{k}\,.
 \end{eqnarray}
Apart from the exceptional case ${\omega=0=\gamma}$, it is
always possible to choose the constants $k_1$ and $k_2$ in such
a way that ${k_1\gamma^2+k_2\,\omega=1}$.

After the transformation (\ref{trans1B}) and the limit
${\kappa\to0}$ are performed, the null tetrad
(\ref{GeneralTetrad}) for the metric (\ref{nonExpMetric}),
(\ref{coeffnonexp}) becomes
 \begin{eqnarray}
 \boldk  \rovno  {\displaystyle \frac{1-\alpha\gamma p}{\sqrt{2}\,\varrho}\left[
 \frac{1}{\sqrt{\cal Q}}\,(2\omega\gamma q\,\partial_{y}-\partial_{t})-\sqrt{\cal Q}\,\partial_{q} \right]} ,  \nonumber\\
 \boldl  \rovno  {\displaystyle \frac{1-\alpha\gamma p}{\sqrt{2}\,\varrho}\left[
 \frac{1}{\sqrt {\cal Q}}\,(2\omega\gamma q\,\partial_{y}-\partial_{t})+\sqrt{\cal Q}\,\partial_{q} \right]} ,  \label{Tetradnonexp}\\
 \boldm  \rovno  {\displaystyle \frac{1-\alpha\gamma p}{\sqrt{2}}\,\left[
 -\frac{\varrho}{\sqrt{{\cal P}^\ex}}\,\partial_y +\im\,\frac{\sqrt{{\cal P}^\ex}}{\varrho}\,\partial_p \right]},
 \nonumber
\end{eqnarray}
with (\ref{spincoeffts}) now taking the form ${\rho = 0 = \mu}$
(because ${{\cal Q}^\ex=\kappa^2{\cal Q}\to0}$). The double
degenerate principal null directions $\boldk $ and $\boldl$
given by (\ref{Tetradnonexp}) are therefore
\emph{non-expanding} and \emph{non-twisting}. The curvature
tensor (\ref{Weyl1}) becomes
\begin{equation}
\Psi_2=-(m+\im\,n)\left(\frac{1-\alpha\gamma p}{\gamma+\im\,\omega p}\right)^3
+(e^2+g^2)\left(\frac{1-\alpha\gamma p}{\gamma+\im\,\omega p}\right)^2 \frac{1-\alpha^2\gamma^2 p^2}{\gamma^2+\omega^2p^2}\,,
\label{Psi2nonexpg}
\end{equation}
while the Ricci tensor (\ref{Ricci1}) now reads
\begin{equation}\Phi_{11}= \frac{1}{2}\,(e^2+g^2)\,\frac{(1-\alpha\gamma p)^4}{(\gamma^2+\omega^2p^2)^2}\,. \label{Phi11nonexpg}
\end{equation}

Such solutions contain the \emph{charge parameters} $e$ and
$g$, the \emph{cosmological constant}~$\Lambda$ and \emph{six
additional parameters} ${\alpha, \omega, n, \gamma}$ and
${\epsilon_0, \epsilon_2}$ (entering ${\cal Q}$). The
parameters $k$, $\epsilon$, $m$, which also occur in ${\cal
P}^\ex$, are uniquely determined by the constraints
(\ref{epsilon})--(\ref{k}). Explicit elimination gives
\begin{eqnarray}
 k \rovno  \frac{-(e^2+g^2)-\epsilon_2\gamma^2 +2\alpha\gamma^3n\omega^{-1} +\Lambda\gamma^4}
 {\omega^2-3\alpha^2\gamma^4}\,, \nonumber\\
 \epsilon \rovno  \frac{-\epsilon_2(\omega^2+3\alpha^2\gamma^4)
 +6\alpha\gamma(\omega^2-\alpha^2\gamma^4)n\omega^{-1} -6\alpha^2\gamma^2(e^2+g^2)
 +2\Lambda\gamma^2\omega^2}{\omega^2-3\alpha^2\gamma^4}\,, \label{kepsilonm}\\
 m \rovno  \frac{-\epsilon_2\gamma(\omega^2+\alpha^2\gamma^4)
 +\alpha\gamma^2(3\omega^2-\alpha^2\gamma^4)n\omega^{-1} -4\alpha^2\gamma^3(e^2+g^2)
 +\frac{4}{3}\Lambda\gamma^3\omega^2}{\omega^2-3\alpha^2\gamma^4}\,. \nonumber
\end{eqnarray}
Now, it needs to be determined whether or not the six
parameters ${\alpha, \omega, n, \gamma, \epsilon_0,
\epsilon_2}$ are independent, and then to determine their
geometrical and/or physical meaning.

\section{General solution: Removing the parameters $\alpha$ and $\omega$}
\label{removingalpha}

We will now show that the parameters $\alpha$ and $\omega$ in
the metric (\ref{nonExpMetric}), (\ref{coeffnonexp}) are, in
fact, \emph{redundant}. It is immediately seen from
(\ref{nonExpMetric})--(\ref{k}) that $\alpha$ plays no
role whenever ${\gamma=0}$ (redefining $\epsilon$, $m$, $k$). Moreover, 
 $\alpha$ can be explicitly transformed away for
any value of~$\gamma$, and $\omega$  can be set to 1 (unless ${\omega=0=\alpha}$),  by
applying the substitution
\begin{equation}
p=\frac{\tilde p-\alpha\gamma^3\mu}{\omega^2\mu+\alpha\gamma\,\tilde p}\,, \qquad
 y=\frac{\tilde y}{\mu}\,, \qquad  \hbox{where}\quad \mu^2=\frac{1}{\omega^2+\alpha^2\gamma^4}\,.
\label{transformationofalpha}
\end{equation}
Under this transformation, the metric (\ref{nonExpMetric}) becomes
\begin{equation}
\d s^2= \tilde\varrho^2\Big(-\widetilde{\cal Q}\,\d t^2 +{1\over\widetilde{\cal Q}}\,\d q^2 \Big)
  +\frac{\widetilde{\cal P}}{\tilde\varrho^2} \Big(\d\tilde y+2\tilde\gamma q\,\d t \Big)^2
  +\frac{\tilde\varrho^2}{\widetilde{\cal P}}\,\d\tilde p^2 ,
\end{equation}
\begin{equation}
 \tilde\varrho^2= \tilde p^2+\tilde\gamma^2\,, \qquad
 \widetilde{\cal Q}= \epsilon_0-\epsilon_2\,q^2\,, \qquad
 \widetilde{\cal P}= a_0 +2\tilde n\,\tilde p +a_2\,\tilde p^2 -{\textstyle\frac{1}{3}}\Lambda\,\tilde p^4\,,
\end{equation}
where
\begin{equation}
\widetilde{\cal P}=\mu^2(\omega^2\mu+\alpha\gamma\,\tilde p)^4\,{\cal P}^\ex\,,\qquad\qquad
\tilde\gamma=\gamma\omega\mu\,,
\end{equation}
with
 $$ \begin{array}{l}
  a_0= -(e^2+g^2)-\epsilon_2\tilde\gamma^2+\Lambda\tilde\gamma^4\,,\qquad
  a_2= \epsilon_2-2\Lambda\tilde\gamma^2\,, \\[4pt]
 2\tilde n/\mu = {\displaystyle-2\alpha\gamma^3\frac{3\omega^2-\alpha^2\gamma^4}{\omega^2-3\alpha^2\gamma^4}\,\epsilon_2
 +2\frac{(\omega^2+\alpha^2\gamma^4)^2}{\omega^2-3\alpha^2\gamma^4}\,\frac{n}{\omega} }\\[4pt]
 \hspace{14mm}{\displaystyle-4\alpha\gamma\frac{\omega^2+\alpha^2\gamma^4}{\omega^2-3\alpha^2\gamma^4}\,(e^2+g^2)
 +\frac{8\alpha\gamma^5\omega^2(3\omega^2-\alpha^2\gamma^4)}{3(\omega^2+\alpha^2\gamma^4)(\omega^2-3\alpha^2\gamma^4)}\,\Lambda}\,. \\
  \end{array} $$
By comparing to (\ref{nonExpMetric}), (\ref{coeffnonexp}), it
can now be seen that the above transformation indeed explicitly
sets ${\omega=1}$ and removes the parameter $\alpha$ from the
metric (after an appropriate relabelling of the parameters $m$,
$n$, $k$ and~$\epsilon$). This is analogous to the case an
\emph{apparently} accelerating NUT metric studied
in~\cite{GriPod05} for which the acceleration parameter
$\alpha$ was similarly shown to be redundant. In fact, the two
transformations are remarkably similar (compare equation
(\ref{transformationofalpha}) with equation (22) in
\cite{GriPod05}).

Notice that (for ${e=0=g}$) the parameter $\alpha$ determines a
kind of \emph{formal rotation in the complex plane} spanned of
the parameters ${m+\im\,n}$, yielding ${\tilde m+\im\,\tilde
n}$. This is clearly seen by performing the substitution
(\ref{transformationofalpha}) in the curvature scalar $\Psi_2$
given by (\ref{Psi2nonexpg}):
\begin{equation}
\Psi_2=-(m+\im\,n)\left(\frac{1-\alpha\gamma\, p}{\gamma+\im\,\omega p}\right)^3
=-c^\frac{3}{2}\,
 \frac{(m+\im\,n)}{\,(\tilde\gamma+\im\,\tilde p)^3}
 \,, \qquad  \hbox{where}\quad c=\frac{\omega+\im\,\alpha\gamma^2}{\omega-\im\,\alpha\gamma^2}\,.
\label{Psi2nonexpgtrasf}
\end{equation}
The parameter $c$ depending on $\omega$ and $\alpha\gamma^2$ is
clearly a \emph{complex unit}. Setting ${\alpha=0}$ by
(\ref{transformationofalpha}) is thus accompanied by a
re-parametrization ${\tilde m+\im\,\tilde n=
c^\frac{3}{2}(m+\im\,n)}$, i.e., mixing the ``original'' $m$
and~$n$.

\vspace{4mm} \textbf{To conclude}: The Pleba\'nski--Demia\'nski
class of non-expanding (electro)vacuum space-times with a
cosmological constant can be written, \emph{without loss of
generality}, by setting ${\alpha=0}$ and ${\omega=1}$ in the
metric (\ref{nonExpMetric}), (\ref{coeffnonexp}) as
 \begin{equation}
 \d s^2= \varrho^2\Big(-{\cal Q}\,\d t^2 +\frac{1}{{\cal Q}}\,\d q^2 \Big)
  +\frac{{\cal P}}{\varrho^2} \Big(\d y+2\gamma q\,\d t \Big)^2
  +\frac{\varrho^2}{{\cal P}}\,\d p^2 ,
 \label{nonExpMetric3}
 \end{equation}
 where, using (\ref{kepsilonm}) with ${\alpha=0}$,
 \begin{eqnarray}
 \varrho^2 \rovno  p^2+\gamma^2 \,, \nonumber\\[3pt]
 {\cal Q}(q) \rovno  \epsilon_0-\epsilon_2\,q^2\,, \label{coeffnonexp3}\\[3pt]
 {\cal P}(p) \rovno  \big(-(e^2+g^2)-\epsilon_2\gamma^2 +\Lambda\gamma^4\big) +2n\,p +(\epsilon_2 -2\Lambda\gamma^2)\,p^2
 -{\textstyle \frac{1}{3}}\Lambda\,p^4\,. \nonumber
\end{eqnarray}
The non-zero components of the curvature tensors are ${R=4\Lambda}$ and
\begin{equation}
\Psi_2=\frac{\epsilon_2\gamma-\frac{4}{3}\Lambda\gamma^3-\im\,n}{(\gamma+\im\,p)^3}
+\frac{e^2+g^2}{(p^2+\gamma^2)(\gamma+\im\,p)^2}\,, \qquad
 \Phi_{11}= \frac{e^2+g^2}{2(p^2+\gamma^2)^2}\,.
  \label{nonExpMetric3NP}
\end{equation}
This class of solutions contains \emph{two discrete
parameters} $\epsilon_0$ and $\epsilon_2$ (using the remaining
scaling freedom in $q$ and $t$ they take the possible values
${+1, 0, -1}$) and \emph{five continuous parameters} $n$,
$\gamma$ and $e$, $g$, $\Lambda$. Since $e$ and $g$ denote the
electric and magnetic charges, respectively, and $\Lambda$ is
the cosmological constant, it remains to determine the
geometrical meaning of the parameters $\epsilon_0$ and
$\epsilon_2$ and the physical meaning of the parameters $n$ and
$\gamma$. This will be done in
Sections~\ref{Minkowski}--\ref{deSitter}
and~\ref{B-metrics}--\ref{interp_gamma}, respectively. In the
final Section~\ref{interp_eg} we will discuss the complete
family, including the charges $e$ and $g$.

\vspace{4mm} Let us mention that this class of solutions was
first found (employing different notation for the coordinates
and free parameters) in 1968 by Carter \cite{Carter68} as his
family ${[\tilde B(-)]}$, see equations (12)--(15) therein.
Subsequently, it was obtained and discussed as ``generalized
anti-NUT solution'' by Pleba\'nski, see pages 235--237 of
\cite{Plebanski75}, equations (3.35)--(3.40) of
\cite{Plebanski79}, and equations (8)--(9) of \cite{GarPle82}
by Garc\'{\i}a D\'{\i}az and Pleba\'nski. The vacuum case with
${\Lambda=0}$ is also equivalent to the case~IV of Kinnersley
\cite{Kinnersley69}. The relation between the
Pleba\'nski--Demia\'nski class of doubly aligned type~D
Einstein--Maxwell fields (denoted as ${\cal D}$) and other
algebraically special solutions has been thoroughly summarized
in a recent work \cite{VandenBergh17}.

\newpage
\section{The canonical Kundt form of these space-times}
\label{Kundt}

Since these solutions admit an expansion-free, twist-free and shear-free repeated principal null direction of the Weyl tensor, they belong to the Kundt class. It must be possible to express them in the canonical Kundt form. For the case ${\alpha=0}$, which is (as shown in previous section) general, this was explicitly already done in \cite{GriPod09}. To put the metric (\ref{nonExpMetric3}), (\ref{coeffnonexp3}) into the Kundt form, first perform the transformation
\begin{equation}
 z=p\,, \qquad
 y_k= y+2\gamma\int\frac{q}{\cal Q}\,\d q\,, \qquad
 r=(p^2+\gamma^2)\,q\,, \qquad
 u=t-\int\frac{1}{\cal Q}\,\d q\,,
 \label{transftoKundt}
\end{equation}
which takes the metric to
\begin{equation}
\d s^2 =-2\,\d u\,\d r -2H\,\d u^2 +2W_{y_k}\,\d u\,\d y_k +2W_z\,\d u\,\d z
 +\frac{1}{P^2}\,\d y_k^2 +P^2\,\d z^2\,,
\end{equation}
 with
 $$ \begin{array}{ll}
 {\displaystyle P^2=\frac{(\gamma^2+z^2)}{{\cal P}(z)}\,,} \\[15pt]
 {\displaystyle H= \frac{\epsilon_0}{2}(\gamma^2+z^2)
 -{1\over(\gamma^2+z^2)}\left[ \frac{\epsilon_2}{2}+
 \frac{2\gamma^2}{(\gamma^2+z^2)P^2}\right]r^2\,, } \\[15pt]
 {\displaystyle W_{y_k}=\frac{2\gamma}{(\gamma^2+z^2)P^2}\,r\,,} \qquad
 {\displaystyle W_z=\frac{2z}{(\gamma^2+z^2)}\,r\,.}
 \end{array} $$
Now, replace ${z=p}$ by a new coordinate
\begin{equation}
  x=\int P^2(z)\,\d z\,,
  \label{ztox}
\end{equation}
 which puts the metric to the Kundt (real) form
\begin{equation}
 \d s^2 =-2\,\d u\,\d r -2H\,\d u^2 +2W_x\,\d u\,\d x +2W_{y_k}\,\d u\,\d y_k
 +P^{-2}(\d x^2+\d y_k^2)\,,
\label{Kundtform1}
 \end{equation}
 where
 $$ W_x=\frac{2z}{(\gamma^2+z^2)P^2}\,r\,, \qquad
 W_{y_k}=\frac{2\gamma}{(\gamma^2+z^2)P^2}\,r\,, $$
 and all metric functions must be re-expressed as functions of~$x$ via~$z$.
 It is of interest to note that all metric coefficients are independent of~$y_k$. Thus, these space-times admit the \emph{two Killing vectors} $\partial_u$ and~$\partial_{y_k}$. In view of this symmetry, the metric form (\ref{Kundtform1}) may be the most appropriate to use. Moreover, the presence of the spacelike Killing vector~$\partial_{y_k}$ indicates that these space-times could possess axial symmetry.

By putting ${\zeta=\frac{1}{\sqrt2}(x+\im\,y_k)}$, the metric (\ref{Kundtform1}) is then easily expressed in the familiar canonical complex form
 \begin{equation}
 \d s^2 =-2\,\d u\,\big(\d r +H\,\d u +W\,\d\zeta +\bar W\,\d\bar\zeta\big)
 +2P^{-2}\d\zeta\d\bar\zeta\,,
 \label{Kundtform2}
 \end{equation}
 where ${W\equiv-\frac{1}{\sqrt2}(W_x-\im\,W_{y_k})}$ reads
 \begin{equation}
 W=-\frac{\sqrt2}{(z+\im\,\gamma)P^2}\,r\,, \nonumber
 \end{equation}
 in which $z$ and $P$ are functions of the \emph{real part} of $\zeta$ only, see   (\ref{ztox}).
 These expressions are equivalent to those given in Section~18.6 of the monograph~\cite{GriPod09}.

\newpage

\section{Special case ${\alpha=0}$, ${\omega=0}$:  direct-product geometries}
\label{furthercases}

In this particular case it is possible to apply on
(\ref{PleDemMetric}) with ${\tilde{n}\equiv n\omega^{-1}}$ a
transformation
\begin{equation}
p=\beta+\kappa\,\tilde{p}\,,\qquad r=\gamma+\kappa\,\tilde{q}\,,\qquad \sigma=\kappa^{-1}\,\tilde{\sigma}\,,\qquad \tau=b^2\,\kappa^{-1}\,\tilde{\tau}\,,
\end{equation}
which yields
\begin{equation}
\d s^2=-\frac{b^4\,\tilde{\mathcal{Q}}}{(\gamma+\kappa\tilde{q})^2}\,\d \tilde{\tau}^2+(\gamma+\kappa\tilde{q})^2\Big(\frac{1}{\tilde{\mathcal{Q}}}\,\d \tilde{q}^2
+\tilde{\mathcal{P}}\,\d\tilde{\sigma}^2+\frac{1}{\tilde{\mathcal{P}}}\,\d \tilde{p}^2\Big)\,,
\end{equation}
with
\begin{eqnarray}
\tilde{\mathcal{P}} \rovno
\kappa^{-2}\,{\cal P}^\ex
=a_0+a_1\tilde{p}+a_2\,\tilde{p}^2\,,\nonumber\\
\tilde{\mathcal{Q}} \rovno
\kappa^{-2}\,{\cal Q}^\ex
=b_0+b_1\,\tilde{q}+b_2\,\tilde{q}^2\,,
\end{eqnarray}
and
\begin{eqnarray}
a_2=-\epsilon\,,\hspace{36mm} b_2 \rovno \epsilon-2\Lambda\,\gamma^2\,, \nonumber\\
a_1=2\kappa^{-1}(\tilde{n}-\epsilon\,\beta)\,, \hspace{16.7mm}
b_1 \rovno 2\kappa^{-1}\left(-m+\epsilon\,\gamma-{\textstyle\frac{2}{3}}\Lambda\,\gamma^3\right), \nonumber\\
a_0=\kappa^{-2}(k+2\beta\,\tilde{n}-\epsilon\,\beta^2)\,, \quad
b_0 \rovno \kappa^{-2}\left(e^2+g^2-2m\,\gamma+\epsilon\,\gamma^2-{\textstyle\frac{1}{3}}\Lambda\,\gamma^4\right).
\end{eqnarray}

A non-expanding solution is now obtained by performing the
limit ${\kappa\to 0}$, giving
\begin{equation}
\d s^2=b^2\Big(-Y\,\d \tilde{\tau}^2+\frac{1}{Y}\,\d\tilde{q}^2\Big)
+\gamma^2\Big(X\,\d \tilde{\sigma}^2+\frac{1}{X}\,\d \tilde{p}^2\Big)\,,
\label{direct_product}
\end{equation}
where
\begin{equation}
X(\tilde{p})= a_0+a_1\,\tilde{p}+a_2\,\tilde{p}^2\,,\qquad
Y(\tilde{q}) = \frac{b^2}{\gamma^2}\big(b_0+b_1\,\tilde{q}+b_2\,\tilde{q}^2\big)\,.
\label{direct_product2}
\end{equation}
This metric clearly represents the class of geometries which
are the \emph{direct-product of two 2-spaces of constant
curvature} with signatures $(-,+)$ and $(+,+)$. These are the
algebraic type~D or conformally flat, (electro)vacuum
Bertotti--Robinson, Narai, and Pleba\'{n}ski--Hacyan solutions
(see Chapter~7 in \cite{GriPod09}).

\vspace{4mm} \textbf{To summarize}: Starting from the
Pleba\'nski--Demia\'nski metric (\ref{PleDemMetric}) with the
parameters $\alpha$ and $\omega$ non-vanishing, the only
possible non-expanding limit is the metric
(\ref{nonExpMetric3}), (\ref{coeffnonexp3}). When ${\alpha=0}$
and ${\omega=0}$ a separate procedure leads to the well-known
family of direct-product geometries (\ref{direct_product}),
(\ref{direct_product2}).


\section{The Minkowski background: ${\Lambda,n,\gamma,e,g=0}$}
\label{Minkowski}

To understand the geometrical meaning of the parameters
$\epsilon_0$ and $\epsilon_2$, which take the discreet values
${+1, 0, -1}$, we naturally investigate them in the
``background'' situation when all the other five physical
parameters are set to zero. In such a case it follows from
(\ref{nonExpMetric3NP}) that the metric (\ref{nonExpMetric3}),
(\ref{coeffnonexp3}) reduces just to \emph{flat Minkowski
space}.

\subsection{Minkowski space in Pleba\'nski--Demia\'nski coordinates}
\label{Minkowski_space}

Let us consider the above family of solutions in the flat case
in which the parameters ${\Lambda, n, \gamma}$ and ${e, g}$ are
all set to zero. The metric (\ref{nonExpMetric3}) then becomes
 \begin{equation}
 \d s^2= p^2\Big(\!\!-{\cal Q}\,\d t^2 +\frac{1}{\cal Q}\,\d q^2 \Big)
  +\epsilon_2\,\d y^2 +\frac{1}{\epsilon_2}\,\d p^2 \,,
 \label{MinkMetric1}
 \end{equation}
 where ${{\cal Q}= \epsilon_0-\epsilon_2\,q^2}$. To maintain the correct signature ${(-++\,+)}$, the parameter $\epsilon_2$ must be positive and may be taken to be unity, ${\epsilon_2=1}$. The resulting form of Minkowski space
 \begin{equation}
 \d s^2= -p^2(\epsilon_0-q^2)\,\d t^2 +\frac{p^2}{\epsilon_0-q^2}\,\d q^2
  +\d y^2 +\d p^2 \,,
 \label{MinkMetric11}
 \end{equation}
thus contain just a \emph{single parameter} $\epsilon_0$, which
may be taken to be ${\epsilon_0=+1,0,-1 }$. Now we will discuss
these three  possibilities. They are three different choices of
the $t$-$q$ coordinates (foliations)  which do not change the
curvature of the 2-dimensional Lorentzian subspace. Its
Gaussian curvature is given by ${\epsilon_2=1}$.

\subsubsection{The case ${\epsilon_0=+1}$}

In this case the metric (\ref{MinkMetric11}) has the form
\begin{equation}
 \d s^2= -p^2(1-q^2)\,\d t^2 +\frac{p^2}{1-q^2}\,\d q^2 +\d y^2 +\d p^2\,.
\label{MinkMetric6}
\end{equation}
There exist \emph{Killing horizons at ${q=\pm1}$ corresponding to the vector field} $\partial_t$. Clearly, $q$ is a spacelike coordinate and $t$ is timelike when ${q\in(-1,1)}$. Otherwise $q$ is timelike and $t$ spacelike.

\begin{figure}[ht]
\centerline{\includegraphics[scale=0.55]{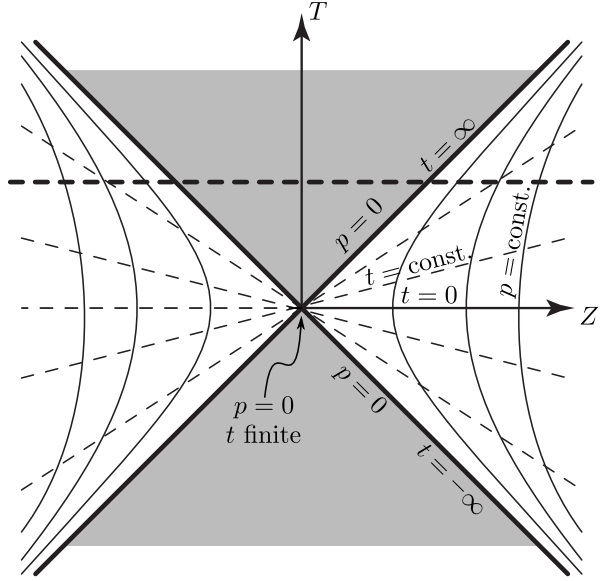}}
\caption{ \small A section of the background flat space on which ${X=0}$ (corresponding to ${q=0}$) and ${Y=y}$ is any constant.
For all three Pleba\'nski--Demia\'nski coordinate parameterisations of Minkowski space with
${\epsilon_0=+1,0,-1}$, the surfaces on which ${p>0}$ is a constant are \emph{rotational hyperboloids}
${-T^2+X^2+Z^2=p^2}$ around the expanding (for ${T>0}$) or contracting (for ${T<0}$) cylinder ${X^2+Z^2=T^2}$, $Y$~arbitrary.
The coordinate singularity ${p=0}$ is located along the $Y$-axis (${T=0}$,
${X=0=Z}$, $Y$~arbitrary). The surfaces on which $t$~is constant are \emph{planes} through the
spacelike line on which ${T=0=Z}$, with $X,Y$ arbitrary. The horizontal heavy dashed line indicates
the section ${T=}$const. through the space-time that is illustrated in Fig.~\ref{e0+1ZXplane}.
The shaded regions are not covered by the Pleba\'nski--Demia\'nski coordinates.}
\label{e0+1ZTplane}
\end{figure}

\begin{figure}[ht]
\centerline{\includegraphics[scale=0.55]{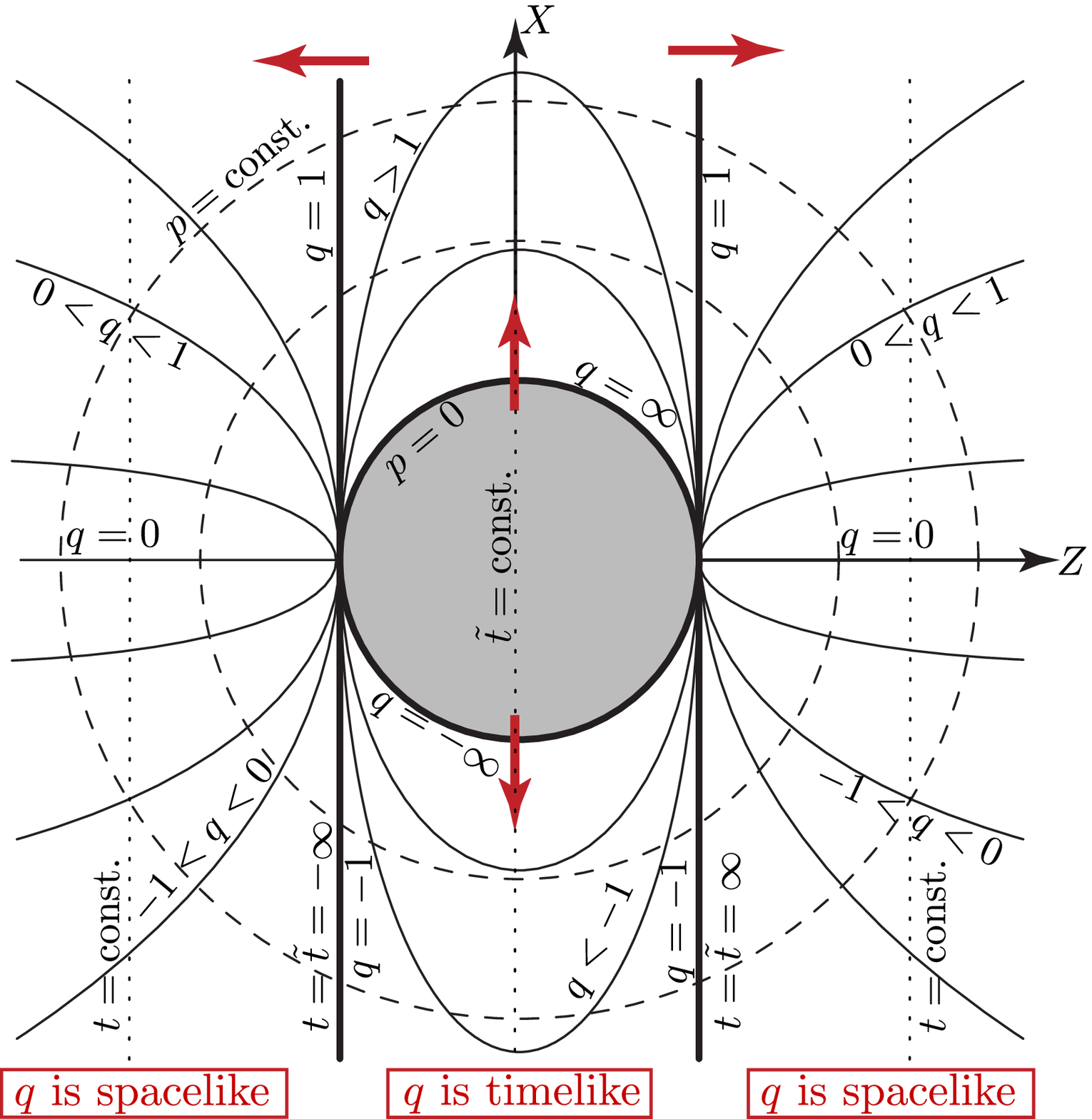}}
\caption{ \small A section of the space-time (\ref{MinkMetric6}) and (\ref{MinkMetric7}) on which $T$ is constant (and $Y$ is arbitrary). For the Pleba\'nski--Demia\'nski
parameterisation of Minkowski space in which ${\epsilon_0=+1}$, the surfaces on which $p$ is a constant are
again rotational hyperboloids (dashed concentric circles in this section) around the expanding/contracting cylinder
${p=0, q=\pm\infty}$, cf. Fig.~\ref{e0+1ZTplane}. Lines on which $q$~is constant are illustrated for the complete range of~$q$ as
hyperbolae (${|q|<1}$) and ellipses (${|q|>1}$). As $T$ increases, the null planes ${q=\pm1}$ (representing
Killing horizons where the norm of the vector field $\partial_t$ vanishes) move apart, and the cylinder
 (whose interior is shaded) on which ${p=0, q=\pm\infty}$ simultaneously contracts/expands, at the speed of light.  }
\label{e0+1ZXplane}
\end{figure}

\vspace{2mm}
$\bullet$ \underline{When ${|q|<1}$}\,, the metric
(\ref{MinkMetric6}) is static, and can be derived from the
usual Cartesian coordinates of Minkowski space
\begin{equation}
 \d s^2= -\d T^2 + \d X^2 +\d Y^2 +\d Z^2
\label{Mink Standard}
\end{equation}
using the transformation
\begin{equation}
\left.
\begin{array}{l}
 T=\pm p\,\sqrt{1-q^2}\sinh t\,, \\[8pt]
 X=p\,q\,, \\[8pt]
 Y=y\,, \\[8pt]
 Z=\pm p\,\sqrt{1-q^2}\cosh t\,,
\end{array}
\right\} \quad\Rightarrow\quad \left\{
\begin{array}{ll}
 p={\displaystyle \sqrt{X^2+Z^2-T^2} }\,, \\[2pt]
 q={\displaystyle \frac{X}{\sqrt{X^2+Z^2-T^2}}}\,, \\[10pt]
 \tanh t={\displaystyle \frac{T}{Z}\,,} \\[6pt]
 y=Y\,,
\end{array}
 \right.
\label{transf1}
\end{equation}
where ${t,y\in(-\infty,\infty)}$ and ${p\in[0,\infty)}$.
Clearly, the surfaces ${p=\,}$const.${\,\not=0}$ and
${q=\,}$const. are geometrically given by
\begin{equation}
-\frac{T^2}{p^2}+\frac{X^2}{p^2} +\frac{Z^2}{p^2}=1 \hbox{\quad and \quad}
\frac{q^2-1}{q^2}\,X^2 +Z^2=T^2\,,
\label{pqsurfaces}
\end{equation}
respectively. The character of these Pleba\'nski--Demia\'nski
coordinates is illustrated in Fig.~\ref{e0+1ZTplane} and
Fig.~\ref{e0+1ZXplane}. The form of the metric
(\ref{MinkMetric6}) is clearly \emph{valid only in the region
${Z^2>T^2}$} outside the pair of null hyperplanes on which
${Z^2=T^2}$ (that is ${t=\pm\infty}$). The \emph{coordinate
singularity} ${p=0}$ for any finite $t$ and $q$ is just the
$Y$-axis, namely  ${T=0}$, ${X=0=Z}$, with $Y$~arbitrary, see
the left part of expression (\ref{transf1}). (We can not use
the inverse relation on the right part of (\ref{transf1}) since
the Jacobian of the transformation is ${|J|=p^2}$, i.e., the
transformation is not regular at ${p=0}$.)

The \emph{Killing horizons} at ${q=\pm1}$ correspond to the two
parts of the null planes ${T=\pm Z}$ with ${X<0}$ for ${q=-1}$,
and ${X>0}$ for ${q=1}$.

Since ${q\in(-1,1)}$, it is natural to put ${q=\cos\theta}$,
${\theta\in(0,\pi)}$, and the metric (\ref{MinkMetric6})
becomes
 \begin{equation}
 \d s^2= p^2(-\sin^2\theta\,\d t^2 +\d\theta^2) +\d y^2 +\d p^2\,.
 \label{MinkMetric4}
 \end{equation}
Interestingly, this form of the metric may be obtained directly from the Cartesian form of Minkowski space (\ref{Mink Standard}) by first applying a Rindler boost
\begin{equation}
\left.
\begin{array}{l}
 T=\tilde z\,\sinh t\,, \\[8pt]
 Z=\tilde z\,\cosh t\,,
\end{array}
\right\} \quad\Rightarrow\quad
\left\{
\begin{array}{l}
\tanh t={\displaystyle \frac{T}{Z}}\,, \\[8pt]
 \tilde z=\sqrt{Z^2-T^2}\,,
\end{array}
\right.
 \label{transf2}
\end{equation}
 in the $Z$-direction, thus giving the metric
\begin{equation}
\d s^2=-\tilde z^2\d t^2 +\d X^2+\d Y^2+\d\tilde z^2\,.
\label{Rindler1}
\end{equation}
By the introduction of standard polar coordinates in the $X,\tilde z$-plane, namely
\begin{equation}
\left.
\begin{array}{l}
 X=p\,\cos\theta\,, \\[10pt]
 \tilde z\ =p\,\sin\theta\,,
\end{array}
\right\} \quad\Rightarrow\quad
\left\{
\begin{array}{ll}
 p={\displaystyle \sqrt{X^2+\tilde z^2} }\,, \\[2pt]
 \tan\theta={\displaystyle \frac{\tilde z}{X}}\,,
\end{array}
\right.
 \label{transf3}
\end{equation}
 and the relabelling ${Y=y}$, we obtain the metric (\ref{MinkMetric4}).
 \pagebreak

\vspace{2mm} $\bullet$ \underline{When ${|q|>1}$}\,, $q$ is a
timelike coordinate while $t$ is spacelike. In this
time-dependent region, the metric (\ref{MinkMetric6}) in the
equivalent form
 \begin{equation}
 \d s^2= -\frac{p^2}{q^2-1}\,\d q^2 +p^2(q^2-1)\,\d t^2 +\d y^2 +\d p^2,
 \label{MinkMetric7}
 \end{equation}
 can be derived from the standard coordinates (\ref{Mink Standard}) of Minkowski space using the transformation
\begin{equation}
\left.
\begin{array}{l}
 T=\pm p\,\sqrt{q^2-1}\,\cosh t\,, \\[8pt]
 X=p\,q\,, \\[8pt]
 Y=y\,, \\[8pt]
 Z=\pm p\,\sqrt{q^2-1}\,\sinh t\,,
\end{array}
\right\}  \quad\Rightarrow\quad
 \left\{
\begin{array}{ll}
 p={\displaystyle \sqrt{X^2+Z^2-T^2} }\,, \\[2pt]
 q={\displaystyle \frac{X}{\sqrt{X^2+Z^2-T^2}}}\,, \\[10pt]
 \tanh t={\displaystyle \frac{Z}{T}}\,, \\[8pt]
 y=Y\,.
\end{array}
\right.
\label{transf4}
\end{equation}
This is very similar to (\ref{transf1}), just interchanging $T$
and $Z$ in the relation for $t$ but, here, $q$ is timelike. The
coordinate singularity at ${p=0}$ with any finite $q$ again
corresponds to the $Y$-axis (that is ${T=0}$, ${X=0=Z}$,
$Y$~arbitrary), while ${p=0, q=\pm\infty}$ is a cylindrical
surface ${X^2+Z^2=T^2}$, any $Y$, which contracts/expands at
the speed of light. The metric (\ref{MinkMetric7}) for
${q\in(1,\infty)}$, however, only covers the region of
Minkowski space for which ${X>0}$ between this cylinder and the
horizon represented by the pair of null hyperplanes on which
${T=\pm Z}$ and ${q=1}$. The equivalent region with ${X<0}$ is
covered by the same metric (\ref{MinkMetric7}) with
${q\in(-\infty,-1)}$. The limits where ${q=\pm1}$ are horizons.

The manifold represented by the metric (\ref{MinkMetric6}) with
the full range ${q\in(-\infty,\infty)}$ thus covers the
\emph{complete} region \emph{outside} the expanding/contracting
cylinder ${X^2+Z^2=T^2}$, with $Y$ arbitrary. The regions
inside the cylinder are excluded. The character of such
Pleba\'nski--Demia\'nski coordinates of Minkowski space is
illustrated in Figs.~\ref{e0+1ZTplane} and~\ref{e0+1ZXplane}.

The surfaces ${p=\,}$const.${\,\not=0}$ and ${q=\,}$const. are
again determined by (\ref{pqsurfaces}). On any constant~$T$ the
lines ${p=\,}$const. are concentric circles ${X^2+Z^2=T^2+p^2}$
while the lines ${q=}$const. are hyperbolae for ${|q|<1}$ and
ellipses for  ${|q|>1}$ (in the limiting cases ${q=0}$ and
${|q|=1}$ these degenerate to straight lines ${X=0}$ and
${Z=\pm T}$, respectively). In particular, at ${T=0}$ all the
curves ${q=}$const. are straight radial lines through the
origin ${X=0=Z}$.

Notice finally that it is possible to get the flat metric (\ref{MinkMetric7}) for ${q^2>1}$ by first obtaining the time-dependent Kasner version of Minkowski space from the Cartesian form (\ref{Mink Standard}) using
\begin{equation}
\left.
\begin{array}{l}
 T=\tilde t\,\cosh z\,, \\[8pt]
 Z=\tilde t\,\sinh z\,,
\end{array}
\right\} \quad\Rightarrow\quad
 \left\{
\begin{array}{ll}
 \tilde t={\displaystyle \sqrt{T^2-Z^2} }\,, \\[2pt]
 \tanh z={\displaystyle \frac{Z}{T}}\,,
\end{array}
\right.
\label{transf5}
\end{equation}
 thus giving
\begin{equation}
\d s^2=-\d\tilde t^2+\d X^2+\d Y^2+\tilde t^2\d z^2\,.
\label{Kasnermetric}
\end{equation}
We can then apply to this a Rindler boost in the $X$-direction, namely
\begin{equation}
 \left.
\begin{array}{l}
 \tilde t\ \>=\pm p\,\sinh\tau\,, \\[6pt]
 X=\pm p\,\cosh\tau\,, \\[6pt]
 Y=y\,,
\end{array}
\right\} \quad\Rightarrow\quad
 \left\{
\begin{array}{ll}
 p={\displaystyle \sqrt{X^2-\tilde t^2} }\,, \\[0pt]
\tanh \tau={\displaystyle \frac{{\tilde t}}{X}}\,, \\[6pt]
 y=Y\,.
\end{array}
\right.
\label{trans6}
\end{equation}
 With this, the metric becomes
\begin{equation}
 \d s^2= p^2(-\d\tau^2 +\sinh^2\tau\,\d z^2) +\d y^2 +\d p^2\,,
 \label{MinkMetric5}
\end{equation}
which is exactly the metric (\ref{MinkMetric7}) with
${q=\cosh\tau}$ and ${t=z}$.

\subsubsection{The case ${\epsilon_0=0}$}

\begin{figure}[ht]
\centerline{\includegraphics[scale=0.55]{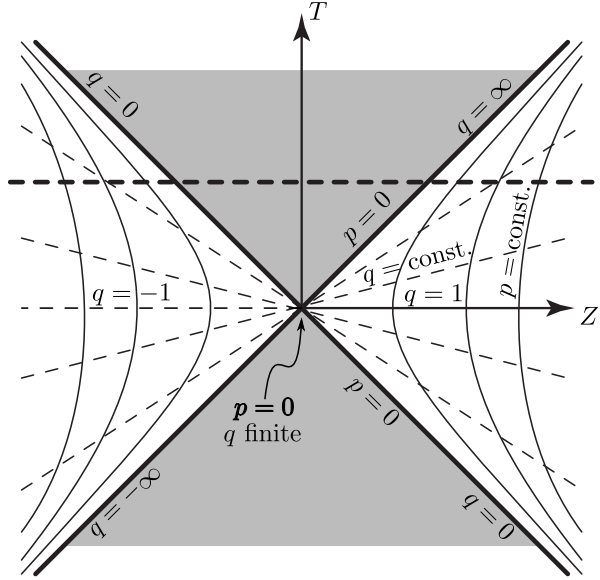}}
\caption{ \small A section ${X=0}$ (corresponding to ${t=0}$) of the background Minkowski space for the
Pleba\'nski--Demia\'nski parameterisation with ${\epsilon_0=0}$. The surfaces on which ${p>0}$ is a
constant are rotational hyperboloids ${-T^2+X^2+Z^2=p^2}$ around the expanding/contracting cylinder
${X^2+Z^2=T^2}$, arbitrary~${Y=y}$, on which ${p=0}$, ${q=\pm\infty}$. The surfaces on which $q$~is constant are
planes through the spacelike line on which ${T=0=Z}$, with $X,Y$ arbitrary.
The horizontal heavy dashed line indicates the section ${T=\,}$const. through the space-time that is
illustrated in Fig.~\ref{e00ZXplane}. The shaded regions are not covered by the Pleba\'nski--Demia\'nski
coordinates.}
\label{e00ZTplane}
\end{figure}

In this case, the Pleba\'nski--Demia\'nski form of the flat metric (\ref{MinkMetric11}) is
\begin{equation}
\d s^2= -\frac{p^2}{q^2}\,\d q^2 +p^2q^2\,\d t^2 +\d y^2 +\d p^2\,.
\label{Mink0}
\end{equation}
It can be derived from the standard form (\ref{Mink Standard}) of Minkowski space via the transformation
\begin{equation}
 \left.
\begin{array}{l}
 T+Z=p\,q\,, \\[6pt]
 T-Z={\displaystyle \frac{p}{q}(q^2t^2-1)}\,, \\[10pt]
 X=p\,q\,t\,, \\[8pt]
 Y=y\,,
\end{array}
\right\} \quad\Rightarrow\quad
 \left\{
\begin{array}{ll}
 p={\displaystyle \sqrt{X^2+Z^2-T^2} }\,, \\[2pt]
 q={\displaystyle \frac{T+Z}{\sqrt{X^2+Z^2-T^2}}}\,, \\[8pt]
 t={\displaystyle \frac{X}{T+Z}}\,, \\[8pt]
 y=Y\,,
\end{array}
\right.
\label{transf7}
\end{equation}
where ${q,t,y\in(-\infty,\infty)}$ and ${p\in[0,\infty)}$. The
coordinate singularity ${p=0}$ (with finite $q, t$) again
corresponds to ${T=0}$, ${X=0=Z}$, $Y$~arbitrary. The surfaces
${p=\,}$const.${\,>0}$ are again given by (\ref{pqsurfaces}),
i.e., they are rotational hyperboloids ${X^2+Z^2=T^2+p^2}$
outside the contracting/expanding cylinder ${X^2+Z^2=T^2}$
(corresponding to the singularity ${p=0}$ with
${q=\pm\infty}$), as shown in Fig.~\ref{e00ZTplane}. For
${X=0}$ this cylinder reduces to ${T=\pm Z}$ which coincides
with the Killing horizon discussed in the case
${\epsilon_0=+1}$.
\begin{figure}[ht]
\centerline{\includegraphics[scale=0.55]{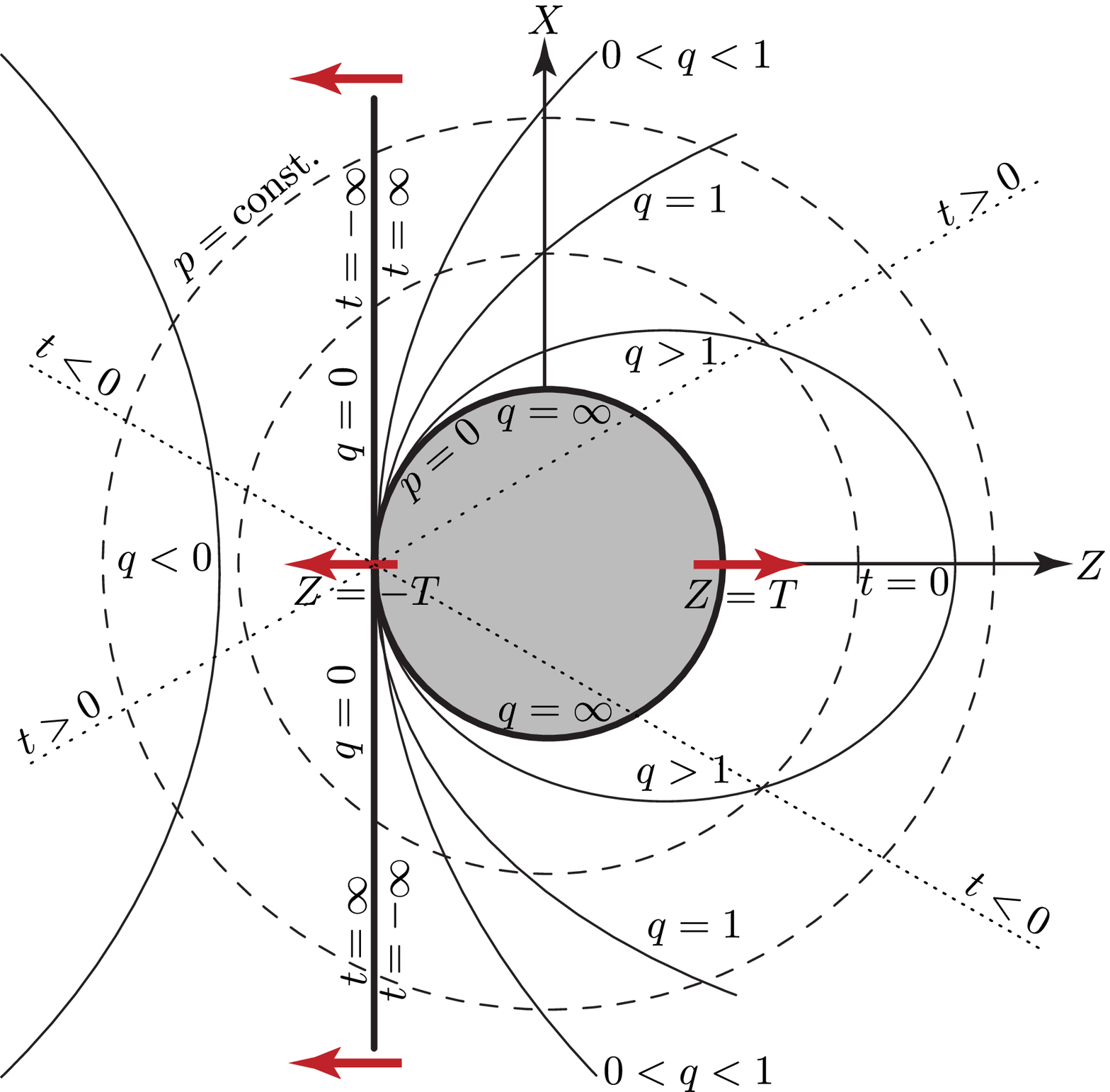}}
\caption{ \small A section of the space-time on which ${T>0}$ is constant (and $Y$ is arbitrary).
For the Pleba\'nski--Demia\'nski parameterisation of Minkowski space in which ${\epsilon_0=0}$,
the surfaces on which $p$ is a constant are again rotational hyperboloids
(dashed concentric circles in this section) around the expanding/contracting  cylinder ${p=0}$.
Lines on which $q$~is constant are illustrated for the complete range: ${q=\infty}$ is a circle
which coincides with the coordinate singularity ${p=0}$, lines ${q>1}$ are ellipses, ${q=1}$ is a parabola, and ${|q|<1}$ are hyperbolae. The coordinate singularity at ${q=0}$ corresponds to the Killing horizon where the norm of $\partial_t$ vanishes. As $T$ increases, the line ${q=0}$ moves to the left and the null cylinder ${p=0}$ expands at the speed of light (see the red arrows).
All the coordinate lines ${q=\,}$const.${\,\ge0}$ intersect in a singular point ${Z=-T}$, which is a
degenerate point on this expanding cylinder (a null line ${X=0}$, ${Z=-T}$, $Y$ arbitrary). }
\label{e00ZXplane}
\end{figure}

However, the surfaces ${q=}$const. and ${t=}$const. are now different, namely
\begin{equation}
X^2 + \frac{q^2-1}{q^2}\bigg(Z-\frac{1}{q^2-1}\,T\bigg)^2=\frac{q^2}{q^2-1}\,T^2\,,
\hbox{\quad and \quad}
X=t\,(T+Z)\,.
\label{pqsurfacesE=0}
\end{equation}
On the section ${X=0}$ this reduces to straight lines ${T=\frac{q^2-1}{q^2+1}\,Z}$, ${T=-Z}$, with ${t=0}$.

On a section on which $T$ is any constant, all the curves
${q=}$const. are \emph{conic sections}. In particular,
${q=\infty}$ corresponds to the circle ${X^2+Z^2=T^2}$ (which
is the singularity ${p=0}$, ${q=\infty}$). For ${q>1}$ the
curves are ellipses with the semi-major axis
${\frac{q^2}{q^2-1}\,T}$ oriented along $Z$. The curve ${q=1}$
degenerates to a parabola ${Z=\frac{1}{2T}X^2-T}$, and for
${|q|<1}$ these coordinate lines are hyperbolae. The line
${q=0}$ is a straight line ${Z=-T}$ with  $X$ arbitrary. This
is illustrated in Fig.~\ref{e00ZXplane}. Moreover, \emph{all
these curves for ${q\ge0}$ intersect at the singular point}
${Z=-T<0}$, ${X=0}$. Notice also that ${q>0\Leftrightarrow
Z>-T}$ whereas ${q<0\Leftrightarrow Z<-T}$. In the special case
${T=0}$ the coordinate lines ${q=\,}$const. are radial straight
lines ${X\propto Z}$. For any fixed~$T$, the coordinate lines
${t=\,}$const. are just \emph{straight lines} ${X=t\,Z+t\,T}$
which all intersect ${X=0}$ at the singular point ${Z=-T}$.

Finally, notice that the flat Pleba\'nski--Demia\'nski-type
metric (\ref{Mink0}) can be rewritten as
\begin{equation}
  \d s^2=p^2(-\d\tau^2+e^{2\tau}\,\d t^2)+\d y^2+\d p^2\,,
\label{Minko0b}
\end{equation}
by introducing ${\tau=\log |q|}$. Clearly, ${p=0}$ is just the
$y$-axis.

\newpage

\subsubsection{The case ${\epsilon_0=-1}$}

In this case, the Pleba\'nski--Demia\'nski form of the metric  (\ref{MinkMetric11}) is
\begin{equation}
 \d s^2= -\frac{p^2}{1+q^2}\,\d q^2 +p^2(1+q^2)\,\d t^2 +\d y^2 +\d p^2\,,
\label{Mink-1}
\end{equation}
in which $q$ is a \emph{timelike} coordinate. This metric can
be derived from the standard Cartesian coordinates of Minkowski
space using the transformation (with the Jacobian ${|J|=p^2}$)
\begin{equation}
 \left.
\begin{array}{l}
 T=p\,q \,, \\[8pt]
 X=p\,\sqrt{1+q^2}\,\sin t \,, \\[8pt]
 Y=y\,, \\[8pt]
 Z=p\,\sqrt{1+q^2}\,\cos t\,,
\end{array}
\right\} \quad\Rightarrow\quad
 \left\{
\begin{array}{ll}
 p={\displaystyle \sqrt{X^2+Z^2-T^2} } \,, \\[4pt]
 q={\displaystyle \frac{T}{\sqrt{X^2+Z^2-T^2}}} \,, \\[10pt]
 \tan t={\displaystyle \frac{X}{Z}} \,, \\[8pt]
 y=Y\,,
\end{array}
\right.
\label{transf8}
\end{equation}
where ${q,y\in(-\infty,\infty)}$ and ${p\in[0,\infty)}$. It can
again be seen that the coordinate singularity at ${p=0}$
corresponds to ${T=0}$, ${X=0=Z}$, with $Y$~arbitrary. The
surfaces ${p=\,}$const.${\,>0}$ are rotational hyperboloids
(\ref{pqsurfaces}) outside the cylinder ${X^2+Z^2=T^2}$ which
expands or contracts at the speed of light (corresponding to
the singularity ${p=0}$ with ${q=\infty}$ or ${q=-\infty}$,
respectively). The above metric only represents the region that
is exterior to this hypersurface.

\begin{figure}[ht]
\centerline{\includegraphics[scale=0.55]{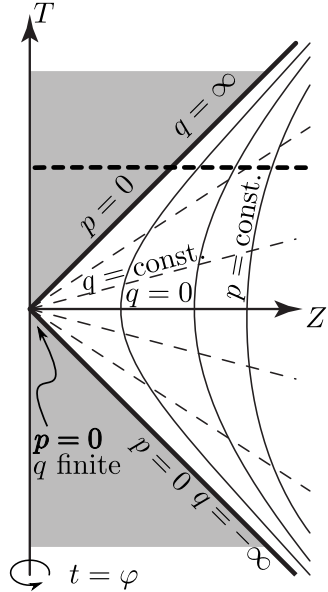}}
\caption{ \small A section of the background flat space with ${\epsilon_0=-1}$
on which ${X=0}$ (and ${Y=y}$ is any constant), corresponding to ${t=0}$.
The surfaces on which ${p>0}$ is a constant are rotational hyperboloids ${-T^2+X^2+Z^2=p^2}$
around the expanding/contracting cylinder ${X^2+Z^2=T^2}$, $Y$~arbitrary, on which ${p=0,\, q=\pm\infty}$.
The surfaces on which $q$~is constant are cones with vertices on the spacelike plane $X,Y$ arbitrary
and ${T=0=Z}$. The shaded regions are not covered. A typical horizontal section ${T=}$~const.
through the space-time is illustrated in Fig.~\ref{e0-1ZXplane}.}
\label{e0-1ZTplane}
\end{figure}

\begin{figure}[ht]
\centerline{\includegraphics[scale=0.55]{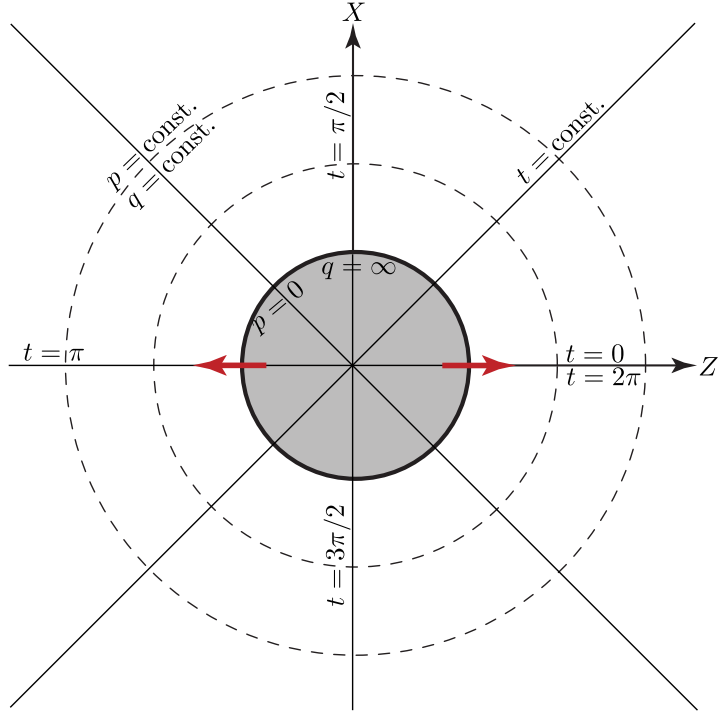}}
\caption{ \small A section ${T=}$~const. through the flat space-time with
${\epsilon_0=-1}$ (${Y=y}$ is constant). Both the lines  ${p=}$~const. and ${q=}$~const. are circles,
while ${t=}$~const. are radial straight lines. In this parameterization (\ref{Mink-1})
there is no Killing horizon associated with ${\partial_t\equiv\partial_\varphi}$.}
\label{e0-1ZXplane}
\end{figure}

It is also now clear from (\ref{transf8}) that the spatial
coordinate $t$ may be taken to be \emph{periodic} with
${t\in[0,2\pi)}$ and ${t=2\pi}$ identified with ${t=0}$. And,
with this angular coordinate $t$, the complete exterior is
covered. This is illustrated in Fig.~\ref{e0-1ZTplane} and
Fig.~\ref{e0-1ZXplane}.

It is thus appropriate to relabel ${t\equiv\varphi}$ and to put
${q=\sinh\tau}$, so that the metric (\ref{Mink-1}) takes the
form
\begin{equation}
 \d s^2=p^2(-\d\tau^2+\cosh^2\tau\,\d\varphi^2)+\d y^2+\d p^2\,.
\label{MinkMetric8}
 \end{equation}
Interestingly, this form of the metric may be obtained from the
Cartesian form of Minkowski space by first introducing polar
coordinates in the $X$-$Z$ plane as
\begin{equation}
 \left.
\begin{array}{l}
 X=\rho\,\sin\varphi\,, \\[8pt]
 Z=\rho\,\cos\varphi\,,
\end{array}
\right\} \quad\Rightarrow\quad
 \left\{
\begin{array}{l}
 \rho=\sqrt{X^2+Z^2}\,, \\[2pt]
 \tan\varphi={\displaystyle \frac{X}{Z}}\,,
\end{array}
\right.
\label{transf9}
\end{equation}
thus giving the cylindrical metric
\begin{equation}
 \d s^2=-\d T^2 +\d\rho^2+\d Y^2+\rho^2\d\varphi^2\,,
 \label{Minkcylindr}
\end{equation}
 and then applying a Rindler boost in the $\rho$-direction, namely
\begin{equation}
 \left.
\begin{array}{ll}
 p=\sqrt{\rho^2-T^2} \,, \\[2pt]
 \tanh \tau={\displaystyle \frac{T}{\rho}}\,,
\end{array}
\right\} \quad\Rightarrow\quad
 \left\{
\begin{array}{l}
 T=p\sinh\tau\,, \\[8pt]
 \rho=p\cosh\tau\,,
\end{array}
\right.
 \label{transf10}
\end{equation}
with ${Y=y}$. The metric (\ref{MinkMetric8}) may thus be
understood as specific accelerating coordinates.

Of course, direct transformations between the metric forms
(\ref{MinkMetric6}), (\ref{Mink0}), and (\ref{Mink-1}) can be
easily obtained by comparing the relations (\ref{transf1}),
(\ref{transf4}), (\ref{transf7}), and (\ref{transf8}).

\newpage

\section{The (anti-)de~Sitter background: ${\Lambda\ne0 \ (n,\gamma,e,g=0)}$}
\label{deSitter}

Consider now the above family of Pleba\'nski--Demia\'nski
solutions in the \emph{conformally flat subcase} in which
$\gamma$, $n$, $e$, $g$ are all set to zero but
${\Lambda\ne0}$, see (\ref{nonExpMetric3NP}). The metric
(\ref{nonExpMetric3}), (\ref{coeffnonexp3}) then reads
\begin{equation}
\d s^2=p^2\Big(-{\cal Q}\,\d t^2 +\frac{1}{{\cal Q}}\,\d q^2 \Big)
+P\,\d y^2 +\frac{1}{P}\,\d p^2\,, \label{PDLambda}
\end{equation}
where ${\cal Q}$ and ${P \equiv {\cal P}/p^2}$ are
 \begin{equation}
{\cal Q}(q)=\epsilon_0-\epsilon_2\,q^2\,, \qquad
 P(p)=\epsilon_2-{\textstyle\frac{1}{3}}\Lambda\, p^2\,.
 \label{PDLambdaPQ}
\end{equation}
This is an \emph{unusual family of metrics of the maximally
symmetric de~Sitter and anti-de~Sitter space-times}. The
(anti-)de~Sitter manifold can be visualized, see
e.g.~\cite{GriPod09}, as the hyperboloid
\begin{equation}
-Z_0^2+Z_1^2+Z_2^2+Z_3^2+\varepsilon Z_4^2=\varepsilon a^2 \,,\quad \hbox{ where }\quad a=\sqrt{3/|\Lambda|}\,,
\quad \varepsilon=\hbox{sign}\,\Lambda \,,
\label{C1}
\end{equation}
embedded in a flat five-dimensional Minkowski space
\begin{equation}
\d s^2=-\d Z_0^2+\d Z_1^2+\d Z_2^2+\d Z_3^2+\varepsilon\d Z_4^2 \,.
\label{C2}
\end{equation}
The coordinates of (\ref{PDLambda}) are adapted to a specific
${2+2}$ foliations of this manifold, and the geometry of such
parametrizations is a \emph{warped product} of two
\emph{2-spaces of constant curvature}, namely ${{dS}_2, {M}_2,
{AdS}_2}$ (according to the sign of $\epsilon_2$) spanned by
${t,q}$\,, and ${{S}^2, {E}^2, {H}^{2}}$ (according to sign of
$\Lambda$) spanned by ${y, p}$. The warp factor is $p^2$.

In our recent work \cite{PodHru17} we have thoroughly studied
and visualized this new family of diagonal static metrics for
all possible choices of $\epsilon_0$, $\epsilon_2$ and for any
${\Lambda\not=0}$. In fact there are 3 allowed distinct
subcases for ${\Lambda>0}$ and 8 subcases for ${\Lambda<0}$,
summarized in Tab.~\ref{tbl:dS}. It is not necessary to repeat
all the specific metric forms, transformations, figures and
other details presented in \cite{PodHru17}. In this section we
will only mention the most interesting subcases of such
Pleba\'nski--Demia\'nski representation of (anti-)de~Sitter
spaces.

\begin{table}[h]
\begin{center}
\begin{tabular}{| c | c | c || c | c|| c | c |}
\hline
$\Lambda$ & $\epsilon_2$ & $\epsilon_0$ & $P$ & range of $p$ & ${\cal Q}$ & range of $q$ \\
\hline
\hline
$>0$ & +1 & +1   & $1-p^{2}/a^{2}$ & $(-a,a)$ & $1-q^{2}$ & $\mathbb{R}\setminus\{\pm 1\}$ \\
\hline
$>0$ & +1 & 0   & $1-p^{2}/a^{2}$ & $[0,a)$  & $-q^{2}$ & $\mathbb{R}\setminus\{0\}$ \\
\hline
$>0$ & +1 & $-1$& $1-p^{2}/a^{2}$ & $[0,a)$ & $-1-q^{2}$ & $\mathbb{R}$ \\
\hline
\hline
$<0$ & +1 & +1 & $1+p^{2}/a^{2}$ & $\mathbb{R}$ & $1-q^{2}$ & $\mathbb{R}\setminus\{\pm 1\}$ \\
\hline
$<0$ & +1 & 0 & $1+p^{2}/a^{2}$ & $[0,\infty)$ & $-q^{2}$ & $\mathbb{R}\setminus\{0\}$ \\
\hline
$<0$ & +1 & $-1$& $1+p^{2}/a^{2}$ & $[0,\infty)$ & $-1-q^{2}$ & $\mathbb{R}$ \\
\hline
$<0$ & 0 & $ +1$& $p^{2}/a^{2}$ & $\mathbb{R}$ & $ 1$ & $\mathbb{R}$ \\
\hline
$<0$ & 0 & $-1$& $p^{2}/a^{2}$ & $\mathbb{R}$ & $-1$ & $\mathbb{R}$ \\
\hline
$<0$ & $-1$ & +1& $-1+p^{2}/a^{2}$ & $[a,\infty)$ & $1+q^{2}$ & $\mathbb{R}$ \\
\hline
$<0$ & $-1$ & 0& $-1+p^{2}/a^{2}$ & $[a,\infty)$ & $q^{2}$ & $\mathbb{R}\setminus\{0\}$ \\
\hline
$<0$ & $-1$ & $-1$& $-1+p^{2}/a^{2}$ & $\mathbb{R}\setminus(-a,a)$ & $-1+q^{2}$ & $\mathbb{R}\setminus\{\pm 1\}$ \\
\hline
\end{tabular}
\caption{\small Summary of all admitted subcases given by
different values of the discrete parameters $\epsilon_2$,
$\epsilon_0$, for ${\Lambda>0}$ (upper part) and ${\Lambda<0}$
(lower part).} \label{tbl:dS}
\end{center}
\end{table}

\newpage

\subsection{The de~Sitter space in Pleba\'nski--Demia\'nski coordinates}

\subsubsection{${\Lambda>0}$, ${\epsilon_2=+1}$, ${\epsilon_0=-1}$}

\begin{figure}[ht]
\centerline{
\includegraphics[scale=0.55]{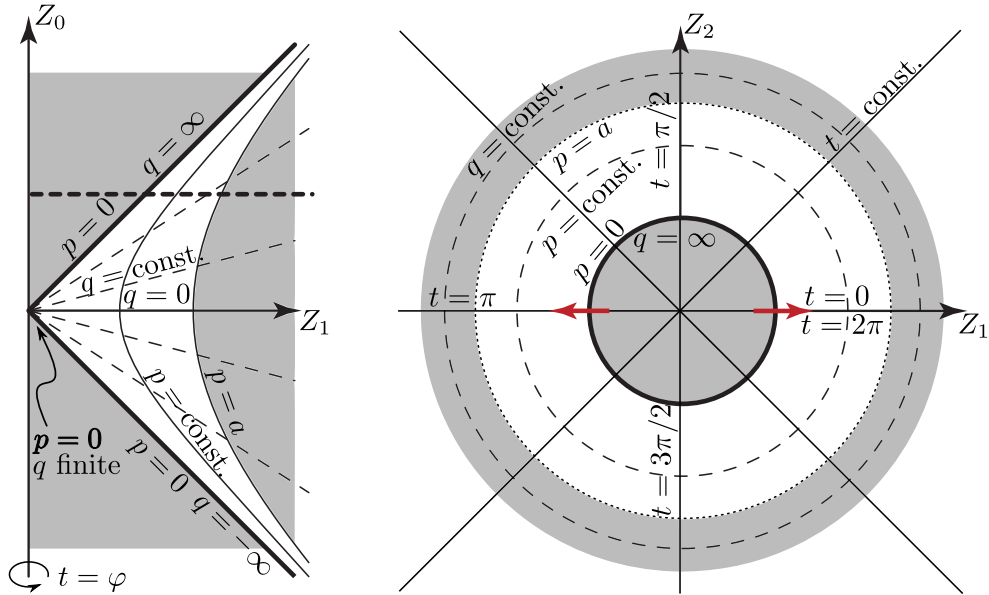}
}
\caption{ \small Sections $Z_1$-$Z_0$ (left) and $Z_1$-$Z_2$ (right)
of the background de~Sitter  space, represented as the hyperboloid (\ref{C1}) in 5-dimensional flat space,
with Pleba\'nski--Demia\'nski coordinates (\ref{PDLambda}), (\ref{PDLambdaPQ})
given by ${\epsilon_2=1}$, ${\epsilon_0=-1}$. The shaded
regions are not covered by these coordinates. Notice a close
similarity with the  ${\epsilon_2=1}$,
${\epsilon_0=-1}$ coordinates of flat Minkowski space
visualized in Fig.~\ref{e0-1ZTplane} and Fig.~\ref{e0-1ZXplane}.}
\label{e+1-1Z0Z1plane_dS}
\end{figure}

This choice of ${\epsilon_2, \epsilon_0}$ seems to be the most
natural one for the case ${\Lambda>0}$. The corresponding
coordinates, with ${y\equiv a\phi}$, cover the (part of) de~Sitter hyperboloid
(\ref{C1}) as

\begin{equation}
\left.
\begin{array}{ll}
 Z_0 = p\,q \,, \\[6pt]
 Z_1 = p\,\sqrt{1+q^2}\,\cos t \,, \\[6pt]
 Z_2 = p\,\sqrt{1+q^2}\,\sin t\,, \\[6pt]
 Z_3 = \sqrt{a^2-p^2}\,\cos \phi\,, \\[6pt]
 Z_4 = \sqrt{a^2-p^2}\,\sin \phi\,,
\end{array}
\right\}
\quad\Rightarrow\quad
\left\{
\begin{array}{l}
 p = \sqrt{Z_{1}^{2}+Z_{2}^{2}-Z_{0}^{2}}\,, \\[2pt]
 q = {\displaystyle \frac{Z_0}{\sqrt{Z_1^2+Z_2^2-Z_0^2}} } \,, \\[12pt]
 \tan t  = {\displaystyle \frac{Z_2}{Z_1}}\,, \\[10pt]
 \tan \phi = {\displaystyle \frac{Z_4}{Z_3}} \,.
 \end{array}
\right. \label{eq:dS1-1par}
\end{equation}
Such a parametrization is visualized in
Fig.~\ref{e+1-1Z0Z1plane_dS} as two sections of the de~Sitter
hyperboloid. The coordinate singularity ${p=0}$ clearly
corresponds to ${Z_0=Z_1=Z_2=0}$, ${Z_3=a\cos\phi}$,
${Z_4=a\sin\phi}$. It is convenient to put ${q=\sinh\tau}$ and
${t=\varphi}$. The metric (\ref{PDLambda}),
(\ref{PDLambdaPQ}) thus takes the form
\begin{equation}
\d s^2= p^2(-\d\tau^2+\cosh^2\tau\,\d\varphi^2) +(a^2-p^2)\,\d \phi^2
+\frac{a^2\,\d p^2}{a^2-p^2}\,.
\label{Rindler2}
\end{equation}
The range of $p$ is finite, namely
${p\in[0,\sqrt{3/\Lambda}\,)}$, to maintain the correct
signature ${(-++\,+)}$, while ${\tau\in\mathbb{R}}$ and
${\varphi, \phi\in[0,2\pi)}$. For ${\Lambda\to 0}$, this
de~Sitter metric reduces to the line element (\ref{MinkMetric8}) of flat
space.

\subsection{The anti-de~Sitter space in Pleba\'nski--Demia\'nski coordinates}

\subsubsection{${\Lambda<0}$, ${\epsilon_2=0}$, ${\epsilon_0=+1}$}
In this case (\ref{PDLambda}) simplifies
considerably to
\begin{equation}
\d s^2=p^2(-\d t^2 +\d q^2 )+\frac{p^2}{a^2}\,\d y^2 +\frac{a^2}{p^2}\,\d p^2\,, \label{PDLambda01}
\end{equation}
where ${a=\sqrt{3/|\Lambda|}}$. With a simple transformation
\begin{equation}\label{relads01}
p = \frac{a^{2}}{x}\,,\qquad
t = \frac{\eta}{a}\,, \qquad
q = \frac{z}{a}\,,
\end{equation}
${\eta,x,y,z\in\mathbb{R}}$ (${x \not=0}$), we obtain the metric
\begin{equation}\label{eq:adScoflat}
\d s^{2}= \frac{a^2}{x^2}\,(-\d \eta^{2}+\d x^{2}+\d y^{2}+\d z^{2})\,.
\end{equation}
This is exactly the \textit{conformally flat Poincar\'e form of
anti-de~Sitter space-time}, see e.g. metric (5.14)
in~\cite{GriPod09}. These well-known coordinates have been
thoroughly described and employed in literature (for example in
the works on AdS/CFT correspondence). The corresponding
explicit parametrization of the anti-de~Sitter hyperboloid
(\ref{C1}) by (\ref{PDLambda01}) is
\begin{figure}[h]
\centerline{
\includegraphics[scale=0.52]{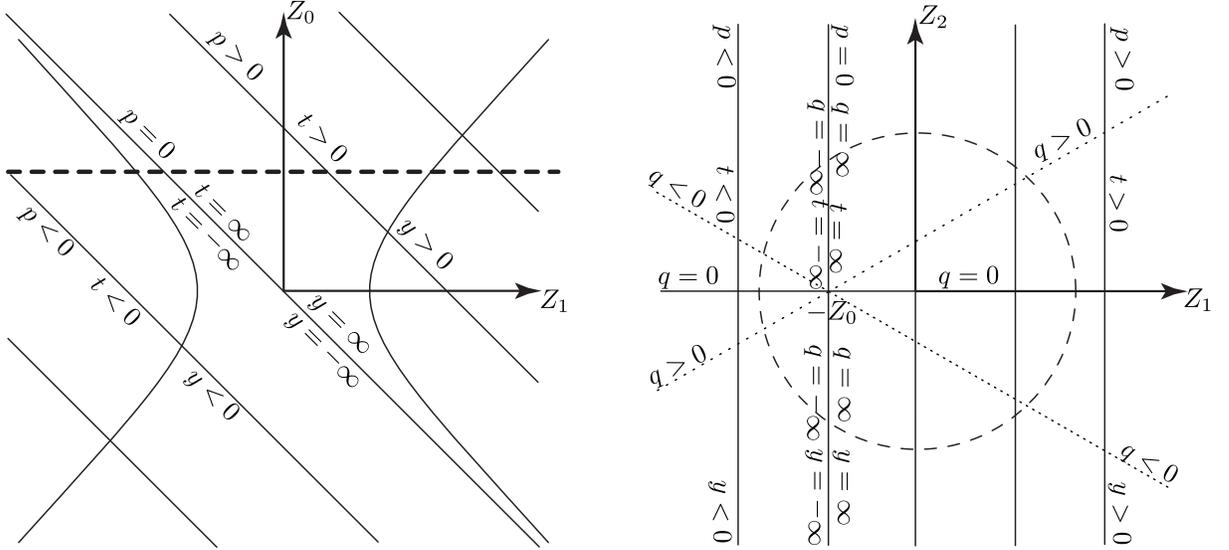}
}
\caption{ \small Sections $Z_1$-$Z_0$ (left) and $Z_1$-$Z_2$ (right) with ${Z_3, Z_4=\hbox{const.}>0}$
of the background anti-de~Sitter  space, represented as the hyperboloid (\ref{C1}) in 5-dimensional flat space,
with Pleba\'nski--Demia\'nski coordinates (\ref{PDLambda01})
given by ${\epsilon_2=0}$, ${\epsilon_0=+1}$.}
\label{e0+1Z0Z1plane_adS}
\end{figure}
\begin{equation}
\left.
\begin{array}{ll}
 Z_0 = {\displaystyle \frac{p}{2}\left(1+\frac{s}{a^2}\right)} \,, \\[6pt]
 Z_1 = {\displaystyle \frac{p}{2}\left(1-\frac{s}{a^2}\right)} \,, \\[3pt]
 Z_2 = p\,q  \,, \\[4pt]
 Z_3 = p\,y/a\,, \\[4pt]
 Z_4 = p\,t  \,,
\end{array}
\right\} \quad\Rightarrow\quad
\left\{
\begin{array}{l}
 p = Z_0+Z_1 \,, \\[4pt]
 q = {\displaystyle \frac{Z_2}{Z_0+Z_1} }\,,\\[8pt]
 t  = {\displaystyle \frac{Z_4}{Z_0+Z_1}}\,, \\[8pt]
 y = {\displaystyle \frac{ aZ_3}{Z_0+Z_1}}\,,
\end{array}
\right.
\label{eq:adS01par}
\end{equation}
where ${s/a^2=-t^2+q^2+y^2/a^2+a^2/p^2}$. The corresponding
sections through the anti-de~Sitter hyperboloids are shown in
Fig.~\ref{e0+1Z0Z1plane_adS}.

\subsubsection{${\Lambda<0}$, ${\epsilon_2=+1}$, ${\epsilon_0=+1}$}
This choice of parameters gives the anti-de~Sitter space in the
Pleba\'nski--Demia\'nski form
\begin{equation}
\d s^2=-p^2(1-q^2)\,\d t^2 +\frac{p^2}{1-q^2}\,\d q^2
+(a^2+p^2)\,\frac{\d y^2}{a^2} +\frac{a^2\,\d p^2}{a^2+p^2}\,,
\label{PDLambda11}
\end{equation}
where ${\,p, t, y \in \mathbb{R}}$,
${q\in\mathbb{R}\setminus\{\pm 1\}}$. This is a generalization
of the flat metric (\ref{MinkMetric6}) to ${\Lambda<0}$. The separate
subcases ${|q|<1}$ and ${|q|>1}$ are:

\vspace{2mm} $\bullet$ \underline{For ${|q|<1}$}\,, the
coordinates of (\ref{PDLambda11}) parametrize the
anti-de~Sitter hyperboloid (\ref{C1}) as
 \begin{equation}
\left.
\begin{array}{l}
Z_0 = {\displaystyle p\,\sqrt{1-q^{2}}\,\sinh t}\,, \\[8pt]
Z_1 = {\displaystyle p\,\sqrt{1-q^{2}}\,\cosh t}\,, \\[8pt]
Z_2 = {\displaystyle |p|\,q}\,, \\[8pt]
Z_3 = {\displaystyle \pm \sqrt{a^2+p^{2}}\,\sinh\frac{y}{a}}\,, \\[8pt]
Z_4 = {\displaystyle \pm \sqrt{a^2+p^{2}}\,\cosh\frac{y}{a}}\,,
\end{array}
\right\} \ \Rightarrow \ \left\{ \!
\begin{array}{l}
\tgh t  = {\displaystyle \frac{Z_0}{Z_1}}\,, \\[10pt]
\tgh {\displaystyle \frac{y}{a}} = {\displaystyle \frac{Z_3}{Z_4}}\,, \\[6pt]
p =  {\displaystyle \sign(Z_1)\,\sqrt{Z_1^2+Z_2^2-Z_0^2}}\,, \\[6pt]
q =  {\displaystyle \frac{Z_2}{\sqrt{Z_1^2+Z_2^2-Z_0^2}}}\,.
\end{array}
\right.
 \label{eq:adS11apar}
 \end{equation}
This parametrization gives two maps covering the anti-de~Sitter
manifold, namely the coordinate map ${Z_4\geq a}$ for the
``$+$'' sign, and   ${Z_4\leq a}$ for the ``$-$'' sign (and two
maps ${p>0}$ and ${p<0}$). Moreover, ${q>0}$ corresponds to
${Z_2>0}$, while ${q<0}$ corresponds to ${Z_2<0}$.

\begin{figure}[ht]
\centerline{
\includegraphics[scale=0.55]{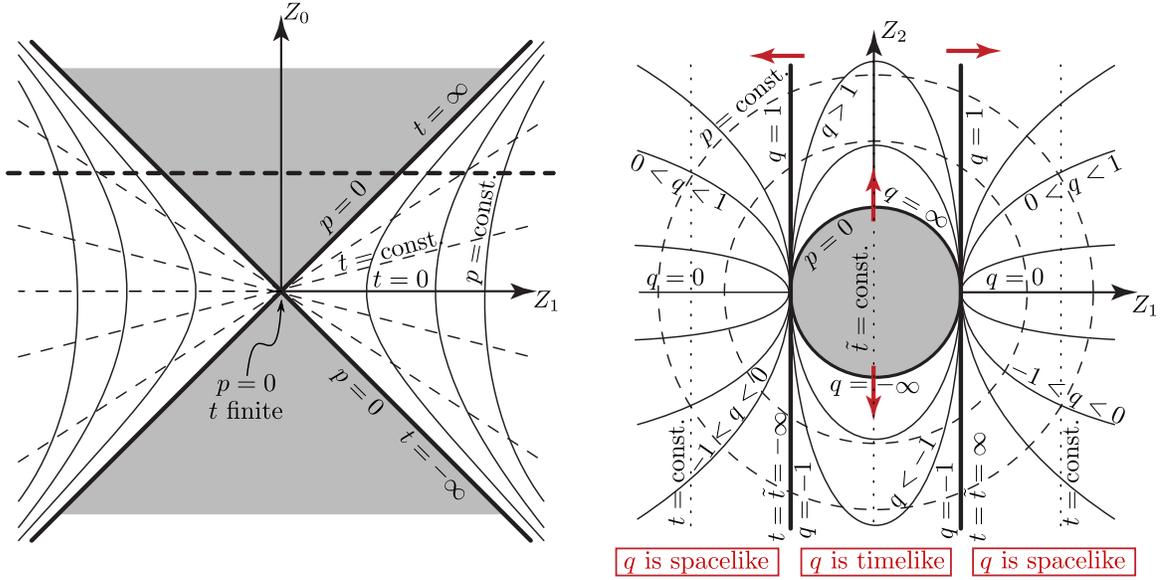}
}
\caption{ \small Sections $Z_1$-$Z_0$ (left) and $Z_1$-$Z_2$ for ${Z_0=\,}$const.${\,>0}$ (right)
of the background anti-de~Sitter  space, represented as the hyperboloid (\ref{C1}) in 5-dimensional flat space,
with Pleba\'nski--Demia\'nski coordinates (\ref{PDLambda11})
given by ${\epsilon_2=+1}$, ${\epsilon_0=+1}$. The shaded
regions are not covered. It resembles the corresponding case of flat Minkowski space
visualized in Fig.~\ref{e0+1ZTplane} and Fig.~\ref{e0+1ZXplane}.}
\label{e+1+1Z0Z1plane_adS}
\end{figure}

\vspace{2mm} $\bullet$ \underline{For ${|q|>1}$}\,, the
parametrization is the same as (\ref{eq:adS11apar}), except
that now
 \begin{equation}
\left.
\begin{array}{l}
Z_0 = {\displaystyle p\,\sqrt{q^{2}-1}\,\cosh t}\,, \\[10pt]
Z_1 = {\displaystyle p\,\sqrt{q^{2}-1}\,\sinh t}\,,
 \end{array}
\right\} \ \Rightarrow \ \left\{ \!
\begin{array}{l}
\tgh t  = {\displaystyle \frac{Z_1}{Z_0}}\,, \\[6pt]
p =  \sign(Z_0)\,\sqrt{Z_1^2+Z_2^2-Z_0^2}\,.
\end{array}
\right.
 \label{eq:adS11bpar}
 \end{equation}

In both cases, it can be immediately observed that the
coordinate \emph{singularity} ${p=0}$ (with finite values of
the coordinates $t, q$) is located at ${Z_0=Z_1=Z_2=0}$ with
${Z_3=\pm a\sinh (y/a)}$, ${Z_4=\pm a\cosh(y/a)}$. This is a
\emph{main hyperbolic line on the hyperboloid} (\ref{C1})
representing the anti-de~Sitter universe.

Sections $Z_1$-$Z_0$ and $Z_1$-$Z_2$ through the anti-de~Sitter
space-time are  illustrated in Fig.~\ref{e+1+1Z0Z1plane_adS}.

\subsubsection{${\Lambda<0}$, ${\epsilon_{2}=-1}$, ${\epsilon_{0}=1}$}
\label{sbs:adS-11}

Relabeling ${y=a\phi}$, the metric (\ref{PDLambda}), (\ref{PDLambdaPQ})  reads
\begin{equation}\label{PDLambda-11}
\d s^{2}=-p^2(1+q^{2})\,\d t^{2}+\frac{p^2}{1+q^2}\,\d q^2
    +(p^2-a^2)\,\d \phi^2+\frac{a^2\,\d p^2}{p^2-a^2}\,,
\end{equation}
where ${p \in [a, \infty)}$, ${q\in\mathbb{R}}$, ${t,\phi\in[0,2\pi)}$, with ${p= a}$ representing
the axis of symmetry.

\begin{figure}[h!]
\centerline{
\includegraphics[scale=0.55]{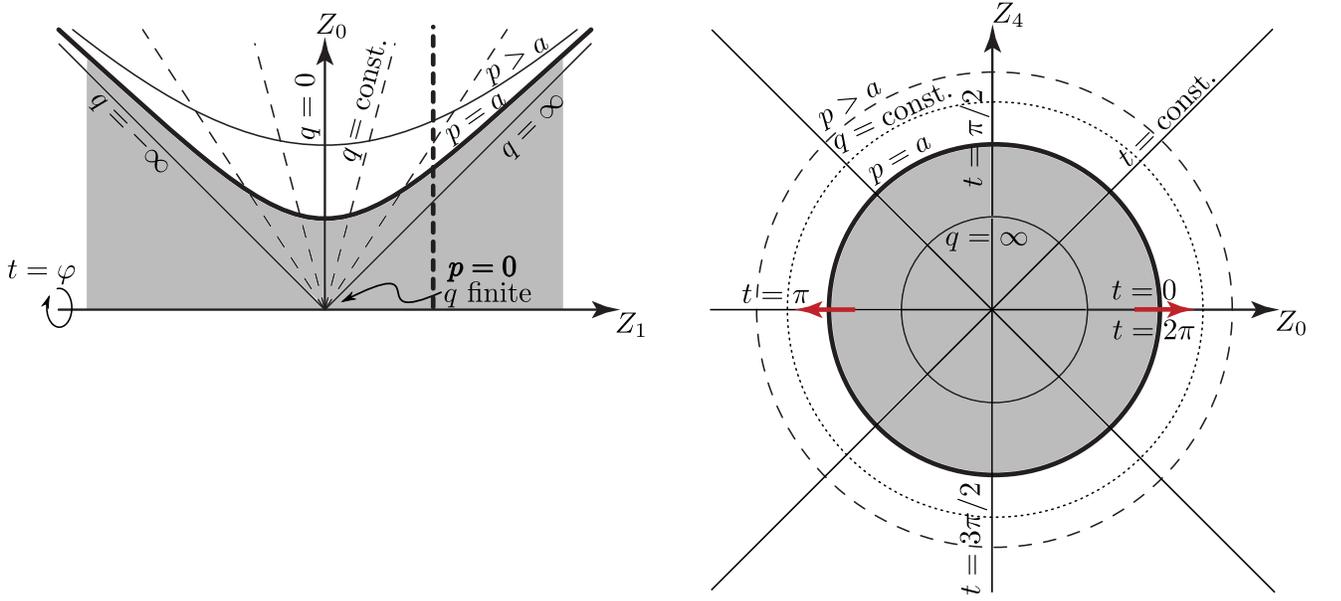}
}
\caption{ \small
Sections $Z_1$-$Z_0$ (left) and $Z_0$-$Z_4$ for ${Z_1=\,}$const.${\,>0}$ (right)
of the anti-de~Sitter  space (\ref{C1})
with Pleba\'nski--Demia\'nski coordinates (\ref{PDLambda-11})
given by ${\epsilon_2=-1}$, ${\epsilon_0=1}$. It resembles the corresponding case of the de Sitter space
visualized in Fig.~\ref{e+1-1Z0Z1plane_dS}.}
\label{e-11Z0Z1plane_adS}
\end{figure}

This arises as the parametrization
\begin{equation}
\left. \begin{array}{ll}
 Z_0 = p\,\sqrt{1+q^2}\,\cos t \,, \\[4pt]
 Z_1 = p\,q \,, \\[3pt]
 Z_2 = \sqrt{p^2-a^2}\,\cos \phi  \,, \\[4pt]
 Z_3 = \sqrt{p^2-a^2}\,\sin \phi  \,, \\[4pt]
 Z_4 = p\,\sqrt{1+q^2}\,\sin t  \,,
 \end{array}\right\} \quad\Leftrightarrow\quad
 \left\{ \begin{array}{l}
 \tan t  = {\displaystyle \frac{Z_4}{Z_0}}\,, \\[8pt]
 \tan \phi = {\displaystyle \frac{Z_3}{Z_2}}\,, \\[8pt]
 p = \sqrt{Z_0^2-Z_1^2+Z_4^{2}}\,, \\[2pt]
 q = {\displaystyle \frac{Z_1}{\sqrt{Z_0^2-Z_1^2+Z_4^2}} }\,,
 \end{array} \right.
\label{eq:adS-11par}
\end{equation}
of the anti-de~Sitter hyperboloid (\ref{C1}). Recall that $Z_0$
a $Z_4$ are two temporal coordinates expressed here by the most
natural single temporal coordinate ${t\in[0,2\pi)}$. The
covering space is obtained by allowing ${t\in\mathbb{R}}$ in
(\ref{PDLambda-11}).

Interestingly, after the formal relabeling $Z_0 \leftrightarrow
Z_1$ and ${Z_2 \rightarrow Z_3 \rightarrow Z_4 \rightarrow
Z_2}$ we obtain basically the same expressions as
(\ref{eq:dS1-1par}) for the de~Sitter subcase ${\Lambda>0}$,
${\epsilon_{2}=1}$, ${\epsilon_{0}=-1}$. Therefore, the
sections through the anti-de~Sitter hyperboloid closely
resemble those shown in Fig.~\ref{e+1-1Z0Z1plane_dS}, after the
relabeling of the axes $Z_a$ and reconsidering different ranges
of the coordinates. In particular, in
Fig.~\ref{e-11Z0Z1plane_adS} we plot the sections ${Z_4=0}$ and
${Z_1=\,}$const.${>0}$, respectively. It can be seen from
(\ref{eq:adS-11par}) that these coordinates \emph{cover the
whole anti-de~Sitter universe}.

More information about the global character of these
coordinates, other cases given by different choices of the
parameters $\epsilon_2$, $\epsilon_0$, their mutual relations and properties
can be found in our previous paper \cite{PodHru17}.

\section{The $B$-metrics: ${n\ne0\ (\gamma,e,g,\Lambda=0)}$}
\label{B-metrics}

To elucidate the meaning of the \emph{physical parameter} $n$,
we first consider the case when ${\gamma=0=\Lambda}$ and
${e=0=g}$. Such vacuum solutions are known as the
$B$\emph{-metrics}, following the classification of Ehlers and
Kundt \cite{EhlersKundt62}.

The subcases of such $B$-metrics are then distinguished by two discrete parameters ${\epsilon_2}$ and~${\epsilon_0}$, with possible values ${+1, 0, -1}$. The corresponding metric (\ref{nonExpMetric3}),
(\ref{coeffnonexp3}) is
\begin{equation}
\d s^2= -p^2(\epsilon_0-\epsilon_2\,q^2)\,\d t^2
 + \frac{p^2}{\epsilon_0-\epsilon_2\,q^2}\,\d q^2
 +\bigg(\epsilon_2+\frac{2n}{p}\bigg)\d y^2
 +\bigg(\epsilon_2+\frac{2n}{p}\bigg)^{-1}\d p^2\,.
\label{Bmetric}
\end{equation}
When $n$ is set to zero, this metric immediately reduces to
background (\ref{MinkMetric1}). The space-times (\ref{Bmetric})
admit four Killing vectors.

In all the subcases, the only non-zero component of the Weyl tensor (\ref{nonExpMetric3NP}) is given by
\begin{equation}
\Psi_2=\frac{n}{p^3}\,,
\label{BmetricPsi2}
\end{equation}
where the two (double degenerate) principal null directions are
 \begin{eqnarray}
 \boldk  \rovno  {\displaystyle -\frac{1}{\sqrt{2}\,p}\Big(
 \frac{1}{\sqrt{\epsilon_0-\epsilon_2\,q^2}}\,\partial_{t}+\sqrt{\epsilon_0-\epsilon_2\,q^2}\,\partial_{q} \Big)} \,,  \nonumber\\
 \boldl  \rovno  {\displaystyle -\frac{1}{\sqrt{2}\,p}\Big(
 \frac{1}{\sqrt {\epsilon_0-\epsilon_2\,q^2}}\,\partial_{t}-\sqrt{\epsilon_0-\epsilon_2\,q^2}\,\partial_{q} \Big)} \,,  \nonumber
\end{eqnarray}
see  (\ref{Tetradnonexp}). They span 2-dimensional spatial
surfaces ${p=}$~const., ${y=}$~const. This confirms that all
$B$-metrics are of \emph{type~D} and possess a \emph{curvature
singularity at} ${p=0}$. It is convenient to consider only
solutions for which $p$ is positive, but the parameter $n$ may
have either sign (notice that the metric only depends on their
fraction $n/p$).

It seems that most of the $B$-metrics have not yet been studied
and physically interpreted, although they are a very simple
family of type~D space-times that have been known for a long
time. Moreover, they are formally related to the well-known
$A$-metrics\index{A-metrics} by a complex coordinate
transformation. If the $y$-coordinate is taken to have a finite
range ${[0,2\pi)}$, with ${y=2\pi}$ identified with ${y=0}$,
the static regions of these space-times can be expressed in
Weyl form. In this case, the associated Newtonian potentials
have been identified by Martins \cite{Martins96} as
semi-infinite line masses. However, the physical interpretation
of these space-times clearly requires further investigation.
Let us present here some observations concerning the physical
and geometrical properties of the class of exact space-times
(\ref{Bmetric}).

\subsection{The $BI$-metric $(\epsilon_2=1)$}

This case ${\epsilon_2=1}$ admits three subcases, namely ${\epsilon_0= 1, 0, -1}$.

\subsubsection{The $BI$-metric with $\epsilon_0=1$}

For the choice ${\epsilon_0=1}$ the line element becomes
 \begin{equation}
 \d s^2= -p^2(1-q^2)\,\d t^2
 + \frac{p^2}{1-q^2}\,\d q^2
 +\bigg(1+\frac{2n}{p}\bigg)\d y^2
 +\bigg(1+\frac{2n}{p}\bigg)^{-1}\d p^2\,,
  \label{BImetric}
 \end{equation}
generalizing (\ref{MinkMetric6}). If ${n>0}$ and
${p\in(0,\infty)}$, there is a \emph{physical singularity at}
${p=0}$. Alternatively, if ${n<0}$, this represents a
\emph{non-singular} space-time with ${p\in(2|n|,\infty)}$. Both
cases are asymptotically flat as ${p\to\infty}$.

The character of the singularity at ${p=0}$ can be elucidated
by considering the weak-field limit of the metric
(\ref{BImetric}) as ${n\to 0}$, with positive $n$. In view of
both explicit transformations (\ref{transf1}) and
(\ref{transf4}) to usual Cartesian coordinates of the
background Minkowski space it is clear, that the
\emph{curvature singularity} at ${p=0}$ \emph{corresponds to}
\begin{equation}
T=0\,, \qquad X=0=Z\,, \qquad Y=y\,.
  \label{tachyonsource}
 \end{equation}
It is localized \emph{along the spatial} $Y$-axis, i.e., it can
be interpreted as the source associated with a tachyon which
moves (with infinite velocity) at ${T=0}$ along the $Y$-axis.
The curved $BI$-metric (\ref{BImetric}) can thus be understood
to include the \emph{gravitational field generated by a tachyon
moving instantaneously along a straight line} (the $y$-axis).

Following an analogy with the $AI$-metric, which represents the
gravitational filed of a static (standing) mass source, it is
natural to put ${q=\cos\theta}$ in (\ref{BImetric}). However,
this is unnecessarily restrictive. The coordinate~$q$ may cover
the complete range ${q\in(-\infty,\infty)}$. For the range
${|q|>1}$, the space-time is time-dependent and $q$~is a
timelike coordinate. Horizons exist at ${q=\pm1}$, between
which the space-time is static. However, from an analysis of
the Minkowski limit as ${n\to0}$, performed in
Section~\ref{Minkowski_space}, it would appear that the Killing
horizons at ${q=\pm1}$ have the character of acceleration
horizons in this particular coordinate representation.
Moreover, this choice of ${\epsilon_0=+1}$ seems to correspond
to an ``unfortunate'' coordinate foliation with coordinate
singularities at ${q=\pm1}$.

\subsubsection{The $BI$-metric with $\epsilon_0=-1$}

Another metric form, which covers the space-time without the
coordinate singularity at ${q=\pm 1}$, occurs with the choice
${\epsilon_0=-1}$. In this case, relabelling
${t=\varphi\in[0,2\pi)}$, the metric is
\begin{equation}
\d s^2= -\frac{p^2}{1+q^2}\,\d q^2 +p^2(1+q^2)\,\d\varphi^2
 +\bigg(1+\frac{2n}{p}\bigg)\d y^2 +\bigg(1+\frac{2n}{p}\bigg)^{-1}\d p^2\,,
 \end{equation}
 generalizing (\ref{Mink-1}). Putting ${q=\sinh\tau}$, it takes the form
\begin{equation}
\d s^2=p^2\left(-\d\tau^2+\cosh^2\tau\,\d\varphi^2\right)
 +\bigg(1+\frac{2n}{p}\bigg)\d y^2 +\bigg(1+\frac{2n}{p}\bigg)^{-1}\d p^2\,.
   \label{BImetric2}
\end{equation}
This is a time-dependent, cylindrically symmetric form of the
curved $BI$-metric, generalizing the flat metric
(\ref{MinkMetric8}). In fact, it is the metric (11.22) in the
paper by Pleba\'nski \cite{Plebanski75}. Again, if ${n>0}$ then
${p\in(0,\infty)}$. Alternatively, if ${n<0}$, this represents
a non-singular region of space-time with ${p\in(2|n|,\infty)}$.
Both cases are asymptotically flat as ${p\to\infty}$.

This form of the $BI$-metric solution was analyzed in 1974 by
Gott \cite{Gott74} and interpreted  as part of the space-time
with ${n<0}$ containing a tachyonic matter source (the other
part of the space-time can be extended by the $AII$-metric).
Indeed, by inspecting the explicit transformation
(\ref{transf8}) to the background Cartesian coordinates we
immediately obtain that the singularity at ${p=0}$ in the
weak-field limit ${n\to 0}$ is again located at
(\ref{tachyonsource}), i.e., along the spatial $Y$-axis. This
confirms that the source of the curvature is a tachyon moving
with infinite velocity along a straight line, namely the
$y$-axis of (\ref{BImetric2}).

\subsubsection{The $BI$-metric with $\epsilon_0=0$}
For the choice ${\epsilon_0=0}$ the $BI$-metric (\ref{Bmetric})
becomes
 \begin{equation}
 \d s^2= - \frac{p^2}{q^2}\,\d q^2 + p^2q^2\,\d t^2
 +\bigg(1+\frac{2n}{p}\bigg)\d y^2
 +\bigg(1+\frac{2n}{p}\bigg)^{-1}\d p^2\,,
  \label{BImetric4}
 \end{equation}
which is a non-flat generalization of Minkowski metric
(\ref{Mink0}), to which it reduces for ${n\to0}$. As can be
seen from expressions (\ref{transf7}), the singular source at
${p=0}$ is also located at (\ref{tachyonsource}), and can again
be physically interpreted as a tachyon moving along the
$y$-axis.

\vspace{4mm} \textbf{To summarize}: The $BI$-metric for any
${\epsilon_0}$ represents a space-time which includes the
\emph{gravitational field of a tachyon of ``strength'' $n$,
moving instantaneously along the straight line given by the
$y$-axis}, that is (\ref{tachyonsource}), which corresponds to
the curvature singularity at ${p=0}$. It seems that the most
natural representation of such solution is given by the metric
(\ref{BImetric2}) for the choice ${\epsilon_0=-1}$ because it
most naturally covers the axially symmetric region of the
space-time (see Figures~\ref{e0-1ZTplane} and
\ref{e0-1ZXplane}) and avoids the additional singularities
associated with the coordinate $q$.

\subsection{The $BII$-metric $(\epsilon_2=-1)$}

In this case in which  ${\epsilon_2=-1}$, it is necessary that ${n>0}$, and the metric takes the form
\begin{equation}
  \d s^2= -p^2(\epsilon_0+q^2)\,\d t^2+ \frac{p^2}{\epsilon_0+q^2}\d q^2
 +\bigg(\frac{2n}{p}-1\bigg)\d y^2 +\bigg(\frac{2n}{p}-1\bigg)^{-1}\d p^2\,,
 \label{BIImetricgen}
 \end{equation}
 with ${p\in(0,2n)}$ and ${q\in(-\infty,\infty)}$, where ${p=0}$ corresponds to a curvature
 singularity and ${p=2n}$ is some kind of pole. Notice that, significantly,
 \emph{this metric does not admit a Minkowski limit as} ${n\to0}$ because this would lead to a wrong signature ${(-+--)}$.

For the choice ${\epsilon_0=-1}$, the metric becomes
 \begin{equation}
 \d s^2= -\frac{p^2}{1-q^2}\,\d q^2  +p^2(1-q^2)\,\d t^2
 +\bigg(\frac{2n}{p}-1\bigg)\d y^2 +\bigg(\frac{2n}{p}-1\bigg)^{-1}\d p^2\,.
 \label{BIImetric}
 \end{equation}
 Again, horizons exist at ${q=\pm1}$, but now the space-time is time-dependent in the range ${|q|<1}$,
 and static elsewhere with temporal coordinate $t$. The additional spatial Killing vector is
$\partial_y$. For the alternative choice ${\epsilon_0=+1}$, the
metric becomes
 \begin{equation}
 \d s^2= -p^2(1+q^2)\,\d t^2+ \frac{p^2}{1+q^2}\d q^2
 +\bigg(\frac{2n}{p}-1\bigg)\d y^2 +\bigg(\frac{2n}{p}-1\bigg)^{-1}\d p^2\,,
 \label{BIImetrica}
 \end{equation}
which is \emph{globally static everywhere}. The same is true
for the choice ${\epsilon_0=0}$, with the metric
 \begin{equation}
 \d s^2= - p^2q^2\,\d t^2 + \frac{p^2}{q^2}\d q^2
 +\bigg(\frac{2n}{p}-1\bigg)\d y^2 +\bigg(\frac{2n}{p}-1\bigg)^{-1}\d p^2\,.
 \label{BIImetricb}
 \end{equation}
The physical meaning of these everywhere curved space-times is,
however, unclear since they do not posses the Minkowski limit
${n \to 0}$.

\subsection{The $BIII$-metric $(\epsilon_2=0)$}

This final case of metric (\ref{Bmetric}) occurs when
${\epsilon_2=0}$,
\begin{equation}
\d s^2= -\epsilon_0\, p^2\,\d t^2
 + \frac{p^2}{\epsilon_0}\,\d q^2
 +\frac{2n}{p}\,\d y^2
 +\frac{p}{2n}\,\d p^2\,.
\label{BIIImetric}
\end{equation}
Necessarily ${\epsilon_0=\pm 1}$, and without loss of
generality we may take ${\epsilon_0=1}$ because the case
${\epsilon_0=-1}$ is equivalent to it via the transformation
${t\leftrightarrow q}$. There is no Minkowski limit ${n\to0}$.
In this case, it is possible to use a remaining scaling freedom
of all coordinates to set ${2n=1}$, and the metric becomes
 \begin{equation}
 \d s^2=p^2( -\d t^2+\d q^2) +\frac{1}{p}\,\d y^2 +p\,\d p^2\,,
 \label{BIIImetrics}
 \end{equation}
which is everywhere static. Performing a simple transformation
\begin{equation}
p=\sqrt\rho\,,\qquad q=C\,\varphi\,,
\end{equation}
we obtain
\begin{equation}
 \d s^2=\rho\,( -\d t^2+C^2\d \varphi^2) +\rho^{-1/2}\,({\textstyle\frac{1}{4}}\d \rho^2 + \d y^2)\,.
\end{equation}
Up to a simple rescaling, this is exactly the \emph{Levi-Civita
solution in the limiting case when} ${\sigma=1/4}$, see
equations (10.11) and (10.14) in \cite{GriPod09}.
Interestingly, this is locally isometric to the asymptotic
form of the Melvin solution, see~(7.21) therein. It can also be
expressed in the form (10.8) with the Kasner-like parameters
${(p_0,p_2,p_3)=(\frac{2}{3},\frac{2}{3},-\frac{1}{3})}$. This
exceptional  space-time is also not yet fully understood
physically.

\section{The $B$-metrics with $\Lambda$: ${n\ne0\ (\gamma,e,g=0)}$}
\label{B-metricsLambda}

The metric (\ref{nonExpMetric3}) now takes the form
 \begin{equation}
 \d s^2= -p^2(\epsilon_0-\epsilon_2\,q^2)\,\d t^2 +\frac{p^2}{\epsilon_0-\epsilon_2\,q^2}\,\d q^2
  +\bigg(\epsilon_2+\frac{2n}{p}-{\textstyle \frac{1}{3}}\Lambda\,p^2\bigg)\d y^2
  +\bigg(\epsilon_2+\frac{2n}{p}-{\textstyle \frac{1}{3}}\Lambda\,p^2\bigg)^{-1}\!\d p^2 \,,
 \label{nonExpMetricBL}
 \end{equation}
 which clearly reduces to the $B$-metric (\ref{Bmetric}) when
 ${\Lambda=0}$ and to (anti-)de~Sitter space in the form (\ref{PDLambda})
 when ${n=0}$.

\subsection{The $BI$-metric with $\Lambda$ $(\epsilon_2=1)$}

Preliminary discussion of this class of exact solutions was
performed in \cite{PodHru17}. It was argued that the $BI$
metrics can be physically interpreted as the
\emph{gravitational field containing a tachyonic source moving
(with infinite velocity) in a de~Sitter or anti-de~Sitter
universe}.

Indeed, by inspecting the representation (\ref{eq:dS1-1par}) of
the de~Sitter hyperboloid for the case ${\Lambda>0}$,
${\epsilon_2=1}$, ${\epsilon_0=-1}$, with ${y=a\phi}$, it
immediately follows that the singularity ${p=0}$ is located at
\begin{equation} Z_0=0\,, \qquad Z_1=0=Z_2\,, \qquad
Z_3=a\cos \phi\,,\qquad Z_4=a\sin \phi\,.
  \label{tachyonsourceDS}
 \end{equation}
In the weak-field limit ${n\to 0}$ this is just the
\emph{``neck'' of the de~Sitter hyperbolid} (\ref{C1}), and it
is a closed circular trajectory of a \emph{spacelike geodesic}
corresponding to a tachyon with an infinite velocity. This
supports the interpretation of the curved $BI$-metric as the
\emph{gravitational field generated by a tachyonic source at
${p=0}$ moving instantaneously around the closed de~Sitter
universe}. In this case it is convenient to put
${q=\sinh\tau}$, ${t=\varphi\in[0,2\pi)}$, so that the metric
(\ref{nonExpMetricBL}) becomes
\begin{equation}
\d s^2= p^2(-\d\tau^2+\cosh^2\tau\,\d\varphi^2) +(1+2n/p
-{\textstyle{\frac{1}{3}}}\Lambda\, p^2)\,a^2\d \phi^2
+\frac{\d p^2}{1+2n/p-\frac{1}{3}\Lambda\, p^2}\,.
\label{Rindler1BL}
\end{equation}
This is a type~D generalization of the conformally flat
de~Sitter metric (\ref{Rindler2}). For ${\Lambda=0}$ this
reduces to the $BI$-metric (\ref{BImetric2}). To maintain the
correct signature ${(-++\,+)}$, the range of $p$ is finite,
namely ${p\in[0,p_{\rm max})}$ such that ${1+2n/p_{\rm max}
-{\frac{1}{3}}\Lambda\, p_{\rm max}^2=0}$.

A similar interpretation is valid also for ${\Lambda<0}$. The
difference is that, using (\ref{eq:adS11apar}),
(\ref{eq:adS11bpar}), the singularity ${p=0}$ is now located at
\begin{equation} Z_0=0\,, \qquad Z_1=0=Z_2\,, \qquad
Z_3=\pm a\sinh (y/a)\,,\qquad Z_4=\pm a\cosh(y/a)\,.
  \label{tachyonsourceADS}
 \end{equation}
This is a \emph{main hyperbolic line on the hyperboloid}
(\ref{C1}) \emph{representing the anti-de~Sitter universe}.
Again, it is spacelike geodesic trajectory of an infinitely
fast tachyon moving along a ``straight line'' in the open
hyperbolic universe with ${\Lambda<0}$. The exact curved
solution can be written in the form (\ref{Rindler1BL}) with
${a\phi}$ replaced by ${y\in(-\infty,\infty)}$.

Analogous results could be obtained for other choices of
$\epsilon_0$, using explicit parameterizations of the de~Sitter
and anti-de~Sitter backgrounds presented in the comprehensive
work~\cite{PodHru17}.

\subsection{The $BII$-metric with $\Lambda$ $(\epsilon_2=-1)$}
This family of metrics reads
 \begin{equation}
 \d s^2= -p^2(\epsilon_0+q^2)\,\d t^2 +\frac{p^2}{\epsilon_0+q^2}\,\d q^2
  +\bigg(\frac{2n}{p}-1-{\textstyle \frac{1}{3}}\Lambda\,p^2\bigg)\d y^2
  +\bigg(\frac{2n}{p}-1-{\textstyle \frac{1}{3}}\Lambda\,p^2\bigg)^{-1}\!\d p^2 \,,
 \label{nonExpMetricBL2}
 \end{equation}
which clearly reduces to (\ref{BIImetricgen}) when
${\Lambda=0}$. The metric (\ref{nonExpMetricBL2}) only has the
required signature for the range of $p$ for which
${2n-p-\frac{1}{3}\Lambda\,p^3>0}$. Thus, it does not admit an
(anti-)de~Sitter limit as ${n\to0}$. This peculiar family of
exact solutions has no obvious physical meaning, unless
${\Lambda<0}$ with $|\Lambda|$ large enough, in which case for
${n=0}$ we obtain the anti-de~Sitter
background~(\ref{PDLambda-11}).

\subsection{The $BIII$-metric with $\Lambda$ $(\epsilon_2=0)$}
In this last case the metric (\ref{nonExpMetricBL}) reduces to
(without loss of generality we may set ${\epsilon_0=1}$)
 \begin{equation}
 \d s^2= p^2(-\d t^2 + \d q^2)
  +\bigg(\frac{2n}{p}-{\textstyle \frac{1}{3}}\Lambda\,p^2\bigg)\d y^2
  +\bigg(\frac{2n}{p}-{\textstyle \frac{1}{3}}\Lambda\,p^2\bigg)^{-1}\!\d p^2 \,,
 \label{nonExpMetric3BL}
 \end{equation}
generalizing (\ref{BIIImetric}).

For ${\Lambda>0}$, ${n>0}$ it is possible to apply the
transformation
\begin{equation}\label{eq:BIIILT1}
{\textstyle
p=\Big(\frac{6n}{\Lambda}\,\sin^2 \frac{\sqrt{3\Lambda}}{2}\rho\Big)^{\frac{1}{3}}\,,\quad
 y=B\left(\frac{3}{4n}\right)^{\frac{1}{3}}\varphi\,,\quad
t=\left(\frac{2}{9n}\right)^{\frac{1}{3}}\tilde{t}\,,\quad
q=C\left(\frac{2}{9n}\right)^{\frac{1}{3}}\tilde{y}\,,}
\end{equation}
obtaining
\begin{equation}
\d s^2={\textstyle\left(\frac{4}{3\Lambda}\sin^2
\frac{\sqrt{3\Lambda}}{2}\rho\right)^{\frac{2}{3}}}
\left(-\d
\tilde{t}^2+C^2\d \tilde{y}^2\right)
+B^2\left({\textstyle\frac{\sqrt{3\Lambda}}{2}}
\frac{\cos^3\frac{\sqrt{3\Lambda}}{2}\rho}
{\sin\frac{\sqrt{3\Lambda}}{2}\rho}\right)^{\frac{2}{3}}\!\d\varphi^2+\d
\rho^2\,.
\end{equation}
This is the \emph{Linet--Tian metric}
\begin{eqnarray}\label{eq:LT}
&&\d s^2=Q^{2/3}\Big(-P^{-2\left(1-8\sigma+4\sigma^2\right)/3\Sigma}\,\d\tilde{t}^2+B^2\,P^{-2\left(1+4\sigma-8\sigma^2\right)/3\Sigma}\,\d\varphi^2\nonumber\\
&&\hspace{24mm} +C^2\,P^{4\left(1-2\sigma-2\sigma^2\right)/3\Sigma}\,\d \tilde{y}^2\Big)+\d\rho^2\,,
\end{eqnarray}
where
\begin{equation}\label{eq:LTdsQP}
{\textstyle
Q(\rho)=\frac{1}{\sqrt{3\Lambda}}\sin\left(\sqrt{3\Lambda}\,\rho\right)\,,\qquad
P(\rho)=\frac{2}{\sqrt{3\Lambda}}\tan\left(\frac{\sqrt{3\Lambda}}{2}\,\rho\right)\,,
}
\end{equation}
see \cite{Linet},\cite{Tian} and \cite{GPLT}, in the case
${\sigma=1/4}$ ($B$ and $C$ are conicity parameters). The
Linet--Tian metric is a generalization of the Levi-Civita
metric to ${\Lambda\not=0}$. It is a \emph{static,
cylindrically symmetric vacuum metric}. The parameter $\sigma$
can be interpreted as the mass density of the source along the
axis ${\rho=0}$.

Alternatively, we can also perform the transformation
\begin{equation}\label{eq:BIIILT2}
{\textstyle
p=\Big(\frac{6n}{\Lambda}\,\cos^2 \frac{\sqrt{3\Lambda}}{2}\rho\Big)^{\frac{1}{3}}\,,\quad
 y=C\left(\frac{4}{3n\Lambda^2}\right)^{\frac{1}{3}} \tilde{y}\,,\quad
t=\left(\frac{\Lambda}{6n}\right)^{\frac{1}{3}}\tilde{t}\,,\quad
q=B\left(\frac{\Lambda}{6n}\right)^{\frac{1}{3}}\varphi\,,}
\end{equation}
leading to
\begin{equation}
\d s^2={\textstyle\cos^\frac{4}{3}\frac{\sqrt{3\Lambda}}{2}\rho}\left(-\d\tilde{t}^2+B^2\,\d\varphi^2\right)
+{\textstyle\frac{4}{3\Lambda}}
\left(
\frac{\sin^3\frac{\sqrt{3\Lambda}}{2}\rho}{\cos\frac{\sqrt{3\Lambda}}{2}\rho}
\right)^{\frac{2}{3}}\!C^2\d \tilde{y}^2
+\d \rho^2\,,
\end{equation}
which is again the metric (\ref{eq:LT}) but now  for
${\sigma=0}$.

In fact, general Linet--Tian metric for ${\Lambda>0}$ is
invariant with respect to a ``duality''
\begin{eqnarray}
\begin{array}{rlrl}
\rho&=\frac{\pi}{\sqrt{3\Lambda}}-\rho'\,, & t&=\big(\frac{4}{3\Lambda}\big)^{(1-8\sigma+4\sigma^2)/\Sigma}\,t'\,,\\
C y&=\big(\frac{4}{3\Lambda}\big)^{-2(1-2\sigma-2\sigma^2)/3\Sigma}\,B'\varphi'\,,\quad &
B\varphi&=\big(\frac{4}{3\Lambda}\big)^{(1+4\sigma-8\sigma^2)/3\Sigma}\,C' y'\,,
\end{array}
\end{eqnarray}
resulting in
\begin{equation}
\sigma=\displaystyle\frac{1-4\sigma'}{4(1-\sigma')}\,.
\end{equation}
For the special choice ${n=\frac{1}{6}\Lambda}$ this relation
between the $BIII$-metric and the Linet--Tian metric can be
found in \cite{GPLT}, but it is clear from (\ref{eq:BIIILT2})
that this transformation exists for any ${n>0}$. Contrary to
the Levi-Civita metric, the Linet--Tian metric does not give
the conformally flat solution when ${\sigma=0}$.

Similar relations apply to ${\Lambda<0}$ with the difference
that, instead of (\ref{eq:LTdsQP}), the functions $P$ and $Q$
 are now
\begin{equation}\label{eq:LTadsQP}
{\textstyle
Q(\rho)=\frac{1}{\sqrt{3|\Lambda|}}\sinh\left(\sqrt{3|\Lambda|}\rho\right)\,,\qquad
P(\rho)=\frac{2}{\sqrt{3|\Lambda|}}\tgh\left(\frac{\sqrt{3|\Lambda|}}{2}\rho\right)\,.
}
\end{equation}
For ${\Lambda<0}$, ${n>0}$ the transformation is
(\ref{eq:BIIILT1}) with
\begin{equation}\label{eq:BIIILT3}
p={\textstyle\bigg(\frac{6n}{|\Lambda|}\,\sinh^2 \frac{\sqrt{3|\Lambda|}}{2}\rho\bigg)^{\frac{1}{3}}}\,,
\end{equation}
yielding the Linet--Tian metric (\ref{eq:LT}) for
${\sigma=1/4}$:
\begin{equation}
\d s^2={\textstyle\left(\frac{4}{3|\Lambda|}\sinh^2 \frac{\sqrt{3|\Lambda|}}{2}\rho\right)^{\frac{2}{3}}}\left(-\d \tilde{t}^2+C^2\d y^2\right)
+B^2\left({\textstyle\frac{\sqrt{3|\Lambda|}}{2}}\frac{\cosh^3 \frac{\sqrt{3|\Lambda|}}{2}\rho}{\sinh \frac{\sqrt{3|\Lambda|}}{2}\rho}\right)^{\frac{2}{3}}\!\d\varphi^2+\d \rho^2\,.
\end{equation}
For ${\Lambda<0}$, ${n<0}$ the transformation is
(\ref{eq:BIIILT2}) with
\begin{equation}
p={\textstyle\bigg(\frac{6n}{\Lambda}\,\cosh^2 \frac{\sqrt{3|\Lambda|}}{2}\rho\bigg)^\frac{1}{3}}\,,
\end{equation}
yielding the Linet--Tian metric for ${\sigma=0}$:
\begin{equation}
\d s^2={\textstyle\cosh^\frac{4}{3}\frac{\sqrt{3|\Lambda|}}{2}\rho}
\left(-\d\tilde{t}^2+B^2\,\d\varphi^2\right)
+{\textstyle\frac{4}{3|\Lambda|}}
\left(\frac{\sinh^3\frac{\sqrt{3|\Lambda|}}{2}\rho} {\cosh\frac{\sqrt{3|\Lambda|}}{2}\rho}\right)^{\frac{2}{3}}\!C^2\d y^2+\d \rho^2\,.
\end{equation}

\textbf{We thus conclude} that the \emph{$BIII$-metrics with
$\Lambda$ are fully equivalent to the Linet--Tian family of
static, cylindrically symmetric metrics with the special value
of ${\sigma=1/4}$, which is dual to ${\sigma=0}$}.

\section{General vacuum case with ${\gamma\ne0\ (e,g=0)}$}
\label{interp_gamma}

Let us now analyze the \emph{most general vacuum metric
 of the non-expanding Pleba\'nski--Demia\'nski class} with any cosmological constant $\Lambda$.
This is easily obtained from (\ref{nonExpMetric3}),
(\ref{coeffnonexp3}) by setting ${e=0=g}$, in which case the
electromagnetic field vanishes:
 \begin{equation}
 \d s^2= \varrho^2\Big(-{\cal Q}\,\d t^2 +\frac{1}{{\cal Q}}\,\d q^2 \Big)
  +\frac{{\cal P}}{\varrho^2} \Big(\d y+2\gamma q\,\d t \Big)^2
  +\frac{\varrho^2}{{\cal P}}\,\d p^2 \,,
 \label{nonExpMetric4}
 \end{equation}
 where
 \begin{eqnarray}
 \varrho^2 \rovno  p^2+\gamma^2\,, \nonumber\\[3pt]
 {\cal Q}(q) \rovno  \epsilon_0-\epsilon_2\,q^2\,, \label{coeffnonexp4}\\[3pt]
 {\cal P}(p) \rovno  \gamma^2(-\epsilon_2+\Lambda\gamma^2) +2n\,p +(\epsilon_2 -2\Lambda\gamma^2)\,p^2
 -{\textstyle \frac{1}{3}}\Lambda\,p^4\,. \nonumber
\end{eqnarray}
This is a generalization of the $B$-metrics (discussed in previous
sections) to include an \emph{additional parameter} $\gamma$.

To clarify the geometrical and physical meaning of this
parameter $\gamma$ we first observe that the corresponding
curvature tensor component (\ref{nonExpMetric3NP}) reduces to
\begin{equation}
\Psi_2=\frac{\gamma\,(\epsilon_2-\frac{4}{3}\Lambda\gamma^2)-\im\,n}{(\gamma+\im\,p)^3}\,.
  \label{nonExpMetric4NP}
\end{equation}
Therefore, the space-times with ${\gamma\ne0}$ are of
\emph{algebraic type D} and contain no curvature singularity.

The only possible exception is
$$ n=0\,,\qquad  {\textstyle\frac{4}{3}\Lambda\gamma^2}=\epsilon_2\,, $$
in which case the space-time is \emph{conformally flat}, with
${ {\cal P} = -\frac{1}{3}\Lambda\,(p^2+\gamma^2)^2}$, that is
 \begin{equation}
 \d s^2= (p^2+\gamma^2)\Big(\!-(\epsilon_0-\epsilon_2\,q^2)\,\d t^2 +\frac{\d q^2}{\epsilon_0-\epsilon_2\,q^2} \Big)
  -{\textstyle \frac{1}{3}}\Lambda\,(p^2+\gamma^2)(\d y+2\gamma q\,\d t )^2
  -\frac{\d p^2}{{\textstyle \frac{1}{3}}\Lambda\,(p^2+\gamma^2)} \,.
 \label{nonExpMetric4ADS}
\end{equation}
Clearly, ${\Lambda \ge 0}$ is prohibited since this would lead
to a degenerate metric or wrong signature. The only possibility
is ${\Lambda<0}$, implying ${\epsilon_2=-1}$ and
${\gamma^2=\frac{1}{4}a^2}$, where ${a=\sqrt{3/|\Lambda|}}$ as
in (\ref{C1}). Being conformally flat and vacuum, such a metric
\begin{equation}
 \d s^2= (p^2+{\textstyle\frac{1}{4}}a^2)\Big(
   \frac{\d q^2}{\epsilon_0+q^2}
   -\epsilon_0\,\d t^2
   +2a^{-1}q\,\d t\,\d y+a^{-2}\d y^2
   \Big)
  +\frac{a^2\,\d p^2}{p^2+\frac{1}{4}a^2}
 \label{nonExpMetric5ADS}
\end{equation}
must be an \emph{unfamiliar metric form of the anti-de~Sitter
space}. Indeed, it is possible to remove the non-diagonal term
$\d t\, \d y$ by performing a linear transformation
\begin{equation}
  t=b_1\,y'+b_2\, t'\,,\qquad
  y=c_1\,y'+c_2\, t'\,,
\end{equation}
where
\begin{equation}
 c_1 =  \sqrt{-\epsilon_0}\,a\,b_1 \,,\qquad\
 c_2 = - \sqrt{-\epsilon_0}\,a\,b_2 \,,
\end{equation}
resulting in
\begin{equation}
 \d s^2= (p^2+{\textstyle\frac{1}{4}}a^2)\Big(\,
 \frac{\d q^2}{\epsilon_0+q^2}
  +2(-\epsilon_0+\sqrt{-\epsilon_0}\,q)\,b_1^2\,\d y'^2
  +2(-\epsilon_0-\sqrt{-\epsilon_0}\,q)\,b_2^2\,\d t'^2
  \Big)
  +\frac{a^2\,\d p^2}{p^2+\frac{1}{4}a^2} \,.
 \label{nonExpMetric5aADS}
\end{equation}
For ${\epsilon_{0}=-1}$ we choose ${b_1 = 1 = b_2}$, implying
${c_1=a=-c_2}$, and the metric becomes
\begin{equation}
 \d s^2= (4p^2+a^2)\Big(\,
   -\frac{{\textstyle\frac{1}{4}}\d q^2}{1-q^2}
  +\frac{1+q}{2}\d y'^2
  +\frac{1-q}{2}\d t'^2
  \Big)
  +\frac{4a^2\,\d p^2}{4p^2+a^2} \,.
 \label{nonExpMetric6ADS}
\end{equation}
When ${q<1}$, a further transformation
\begin{eqnarray}
2p=a\,\sinh\theta\,,\qquad q=\cos 2\chi\,,
\end{eqnarray}
leads to
\begin{eqnarray}\label{adSgammaMetric1}
\d s^2=a^2\cosh^2\theta\left(-\d\chi^2+\cos^2\chi\,\d y'^2+\sin^2\chi\,\d t'^2\right)+a^2\d\theta^2\,.
\end{eqnarray}
This is an interesting \emph{new diagonal metric form of the
anti-de~Sitter space} corresponding to the parametrization
 \begin{equation}
\left.
\begin{array}{l}
Z_0 = a\cosh\theta\cos\chi\cosh y'\,,\\[6pt]
Z_1 = a\cosh\theta\cos\chi\sinh y'\,,\\[6pt]
Z_2 = a\sinh\theta\,,\\[6pt]
Z_3 = a\cosh\theta\sin\chi\sinh t'\,,\\[6pt]
Z_4 = a\cosh\theta\sin\chi\cosh t'\,,\\
\end{array}
\right\} \ \Rightarrow \ \left\{ \!
\begin{array}{l}
\tgh y'  \,= {\displaystyle \frac{Z_1}{Z_0}}\,, \\[10pt]
\tgh t' = {\displaystyle \frac{Z_3}{Z_4}}\,, \\[8pt]
a\sinh\theta = 2p = Z_2\,, \\[2pt]
\tan \chi =  { \sqrt{\frac{Z_4^2-Z_3^2}{Z_0^2-Z_1^2}}}\,,
\end{array}
\right.
 \end{equation}
of the hyperboloid (\ref{C2}).

\noindent
When ${q>1}$, an analogous transformation
\begin{eqnarray}
2p=a\,\sinh\theta\,,\qquad q=\cosh 2\chi\,,
\end{eqnarray}
puts (\ref{nonExpMetric6ADS}) to \emph{another} metric form of
the anti-de~Sitter space
\begin{eqnarray}\label{adSgammaMetric2}
\d s^2=a^2\cosh^2\theta\left(\d\chi^2+\cosh^2\chi\,\d y'^2-\sinh^2\chi\,\d t'^2\right)+a^2\d\theta^2\,,
\end{eqnarray}
corresponding to
 \begin{equation}
\left.
\begin{array}{l}
Z_0 = a\cosh\theta\cosh\chi\cosh y'\,,\\[6pt]
Z_1 = a\cosh\theta\cosh\chi\sinh y'\,,\\[6pt]
Z_2 = a\sinh\theta\,,\\[6pt]
Z_3 = a\cosh\theta\sinh\chi\cosh t'\,,\\[6pt]
Z_4 = a\cosh\theta\sinh\chi\sinh t'\,,\\
\end{array}
\right\} \ \Rightarrow \ \left\{ \!
\begin{array}{l}
\tgh y'  \,= {\displaystyle \frac{Z_1}{Z_0}}\,, \\[10pt]
\tgh t' = {\displaystyle \frac{Z_4}{Z_3}}\,, \\[8pt]
a\sinh\theta = 2p = Z_2\,, \\[2pt]
\tanh \chi =  { \sqrt{\frac{Z_3^2-Z_4^2}{Z_0^2-Z_1^2}}}\,.
\end{array}
\right.
 \end{equation}
In the case when ${\epsilon_{0}=1}$ we choose ${b_1 = \im =
b_2}$, thus ${c_2=a=-c_1}$, and with the reparametrization
${q\to \im \,q}$ the real metric becomes exactly the same as
(\ref{nonExpMetric6ADS}).

\vspace{3mm}

Returning now to type~D space-times (\ref{nonExpMetric4}),
(\ref{coeffnonexp4}) with ${\gamma\ne0}$, it follows from
(\ref{nonExpMetric4NP}) that \emph{they are all non-singular}
and the range of $p$ is ${(-\infty,+\infty)}$. Indeed, the
curvature singularity at ${p=0}$, which is always present when
${\gamma=0}$ (that is for the $B$-metrics described in previous
section, see (\ref{BmetricPsi2})), is not reached since the
denominator ${\gamma+\im\,p}$ in (\ref{nonExpMetric4NP}) now
has a \emph{finite non-zero value} $\gamma$ for ${p=0}$.
Therefore, the privileged value ${p=0}$ does not correspond to
a physical singularity, but rather to a \emph{region of
space-time with a maximum finite curvature} because $\Psi_2$
has the greatest value there.

Such a behavior is analogous to the more familiar situation
known for the $A$-metrics, in particular for the Schwarzschild
solution with the mass parameter $m$. By adding the
parameter~$l$ in the Taub-NUT solution, the curvature
singularity at ${r=0}$ is removed, see e.g. Chapter~12 in
\cite{GriPod09}. Similarly, by adding the parameter $\gamma$ to
the $B$-metrics with a ``tachyonic mass'' parameter~$n$, the
curvature singularity at ${p=0}$ is also removed. This formal
analogy was noticed already by Pleba\'nski in 1975
\cite{Plebanski75} and, because of it, he denoted this class
with ${\gamma\ne0}$ as the so called ``anti-NUT solution''.

However, it  should be noted that there are also some
fundamental geometrical differences between the $A$-metrics and
$B$-metrics case. While for the Taub-NUT solution with the
parameter~$l$ the double degenerate principal null directions
$\boldk $ and $\boldl$ are expanding and twisting, for the
$B$-metrics with the parameter~$\gamma$ they are non-expanding
and non-twisting (in fact, the whole family of solutions
belongs to the Kundt class).

Although there is no curvature singularity at ${p=0}$ when
${\gamma\ne0}$, and \emph{asymptotically the space-times
contain conformally flat regions as} ${p\to\pm\infty}$, see
(\ref{nonExpMetric4NP}) implying ${\Psi_2\to0}$, in general
there are \emph{Killing horizons} associated with the vector
field $\partial_t$. They separate stationary regions from the
dynamical one. Indeed, it follows from the metric form
(\ref{nonExpMetric4}) that the norm of this vector  is
\begin{equation}
 \|\partial_t\|^2 = g_{tt} =  -\varrho^2{\cal Q}+4\gamma^2 q^2\frac{{\cal P}}{\varrho^2} \,,
\label{Killingvector norm}
\end{equation}
where ${\cal Q}$ and ${\cal P}$ are given by (\ref{coeffnonexp4}). The associated Killing horizon is thus located at
\begin{equation}
(p^2+\gamma^2)^2{\cal Q}=4\gamma^2 q^2{\cal P} \,.
\label{Killinghoriozns}
\end{equation}
This is rather complicated expression, polynomial in the
coordinates $q$ and $p$, and also depending on all five
geometrical and physical parameters. Interestingly, in the case
${\gamma=0}$ it simplifies enormously to the condition ${{\cal
Q}=0}$, so that the Killing horizon is simply located at
${q^2=\epsilon_0/\epsilon_2}$. This is basically the same
condition as for the Minkowski and (anti-)de~Sitter backgrounds
discussed in previous Sections~\ref{Minkowski}
and~\ref{deSitter},  respectively, or in~\cite{PodHru17}.

\section{Charged metrics: the most general case with ${e,g\ne0}$}
\label{interp_eg}

It remains to analyze the complete non-expanding
Pleba\'nski--Demia\'nski metric (\ref{nonExpMetric3}),
(\ref{coeffnonexp3}), i.e.,
 \begin{equation}
 \d s^2= (p^2+\gamma^2)\Big(\!\!-(\epsilon_0-\epsilon_2\,q^2)\,\d t^2
 +\frac{\d q^2}{\epsilon_0-\epsilon_2\,q^2} \Big)
  +{\cal R}\,\big(\d y+2\gamma q\,\d t \big)^2
  +\frac{1}{{\cal R}}\,\d p^2\,,
 \label{nonExpMetric3F}
 \end{equation}
 where
 \begin{equation}
{\cal R}(p)=\frac{\big(-(e^2+g^2) -\epsilon_2\gamma^2+\Lambda\gamma^4\big)
+2n\,p +(\epsilon_2 -2\Lambda\gamma^2)\,p^2 -{\textstyle \frac{1}{3}}\Lambda\,p^4}{p^2+\gamma^2}\,.
\label{coeffnonexp3F}
\end{equation}
It contains seven free parameters, namely two discrete
geometric parameters $\epsilon_0$, $\epsilon_2$ and five
physical parameters $\Lambda$, $n$, $\gamma$, plus $e$ and $g$.
When ${e=0=g}$ the space-times are vacuum, as described in
previous sections. For non-vanishing  $e, g$  the Ricci tensor
given by $\Phi_{11}$ is non-zero, see (\ref{nonExpMetric3NP}),
and \emph{such exact space-times contain a non-null
electromagnetic field} (the source-free Maxwell equations are
also satisfied) which is doubly aligned with the repeated null
directions of the gravitational field. In fact, they can be
understood as a large class of $B$-metrics with the ``mass''
parameter $n$, generalized to admit electric a magnetic charges
$e, g$, in addition to the cosmological constant $\Lambda$. As
can be seen from (\ref{nonExpMetric3NP}), with a non-trivial
parameter~$\gamma$, both the gravitational and the
electromagnetic fields are non-singular.

Let us also note that the parameter $\epsilon_0$ is not
physically important. It only distinguishes three coordinate
representations of the 2-space of constant curvature (given by
$\epsilon_2$) spanned by the coordinates $t$ and $q$. For
example, explicit transformation from the
 metric (\ref{nonExpMetric3F}) with
${\epsilon_2=-1}$, ${\epsilon_0=1}$ (with coordinates relabeled
to ${\tilde t, \tilde q, \tilde y}$, ${|\tilde q|<1}$) to the
metric with ${\epsilon_0=-1}$ is
 \begin{equation}
 \tan\tilde t= \frac{-q}{\sqrt{1-q^2}\,\cosh t}\,,\qquad
 \tilde q = \sqrt{1-q^2}\,\sinh t\,,\qquad
 \tilde y = y+2\gamma\arg \tanh(q\tanh t)\,.
 \label{transfwithgamma}
 \end{equation}
Moreover, as shown explicitly for the case ${\epsilon_2=-1}$,
${\epsilon_0=-1}$ in Section~\ref{interp_gamma},  the
off-diagonal term ${\d y\, \d t}$ in such a form of the
anti-de~Sitter background can be completely removed. Similar
arguments  also apply to the case ${\epsilon_0=0}$.

\subsection{The case when ${\gamma=0}$: Charged $B$-metrics with $\Lambda$}

With ${\gamma=0}$, the metric (\ref{nonExpMetric3F}),
(\ref{coeffnonexp3F}) simplifies considerably to
 \begin{equation}
 \d s^2= -p^2(\epsilon_0-\epsilon_2\,q^2)\,\d t^2
  +\frac{p^2}{\epsilon_0-\epsilon_2\,q^2}\,\d q^2
  +{\cal R}\, \d y^2 + \frac{1}{{\cal R}}\,\d p^2 \,,
 \label{nonExpMetric2eg}
 \end{equation}
 where
\begin{equation}
{\cal R}(p) = \epsilon_2 +\frac{2n}{p} - \frac{e^2+g^2}{p^2} -\frac{\Lambda}{3}\, p^2\,.
\end{equation}
The metric functions are now given in terms of two geometrical
parameters  ${\epsilon_0, \epsilon_2 = +1, 0, -1}$, one
``mass'' parameter $n$, the electric/magnetic charge parameters
$e$ and $g$, and the cosmological constant~$\Lambda$.  The
non-zero components of the curvature tensor are
\begin{equation}
\Psi_2=\frac{n}{p^3} -\frac{e^2+g^2}{p^4}\,, \qquad
\Phi_{11}= \frac{e^2+g^2}{2\,p^4}\,.
\end{equation}
Together these clearly
indicate the presence of a \emph{curvature singularity at}
${p=0}$ whenever either $n$ or ${e^2+g^2}$ are non-zero.
Moreover, its ``strength'' is directly proportional to these
parameters. Such solutions represent a generalization of the
$B$-metrics (\ref{Bmetric}), as originally described by Ehlers
and Kundt \cite{EhlersKundt62}, to include charges and a
cosmological constant.

\subsection{The electromagnetic field}

Finally, we will investigate the electromagnetic field
associated with the general space-time (\ref{nonExpMetric3F}),
(\ref{coeffnonexp3F}). It is described by antisymmetric
Faraday--Maxwell tensor $F_{\mu\nu}$, or the related 2-form
\begin{equation}
F={\textstyle\frac{1}{2}}F_{\mu\nu}\,\d x^\mu\wedge\d x^\nu\,.
\end{equation}
Its dual is ${{\tilde{F}_{\mu\nu}\equiv
\frac{1}{2}\varepsilon_{\mu\nu\alpha\beta}F^{\alpha\beta}}}$,
where ${\varepsilon_{0123}=\sqrt{-g}}$. Maxwell's equations
without sources are ${F^{\mu\nu}_{\quad;\nu}=0}$,
${\tilde{F}^{\mu\nu}_{\quad;\nu}=0}$, which can be rewritten as
${\d\om=0}$, where the complex 2-form $\om$ is defined by
${\om\equiv
F+\im\,\tilde{F}=\frac{1}{2}(F_{\mu\nu}+\im\,\tilde{F}_{\mu\nu})\,\d
x^\mu\wedge\d x^\nu}$.

Non-trivial components of the electromagnetic field associated
with (\ref{nonExpMetric3F}), (\ref{coeffnonexp3F}) are
\begin{eqnarray}\label{eq:PDnexF}
F_{q\,t}\rovno-\frac{e(\gamma^2-p^2)+2g\,\gamma\,p}{\gamma^2+p^2}\,,\nonumber\\
F_{ y\,p}\rovno\frac{g(\gamma^2-p^2)-2e\,\gamma\,p}{(\gamma^2+p^2)^2}\,,\\
F_{p\,t}\rovno-2\gamma\,q\,\frac{g(\gamma^2-p^2)-2e\,\gamma\,p}{(\gamma^2+p^2)^2}\,,\nonumber
\end{eqnarray}
and for the dual
\begin{eqnarray}\label{eq:PDnexFt}
\tilde{F}_{q\,t}\rovno-\frac{g(\gamma^2-p^2)-2e\,\gamma\,p}{\gamma^2+p^2}\,,\nonumber\\
\tilde{F}_{ y\,p}\rovno-\frac{e(\gamma^2-p^2)+2g\,\gamma\,p}{(\gamma^2+p^2)^2}\,,\\
\tilde{F}_{p\,t}\rovno2\gamma\,q\,\frac{e(\gamma^2-p^2)+2g\,\gamma\,p}{(\gamma^2+p^2)^2}\,,\nonumber
\end{eqnarray}
see \cite{Hruska2015} for more details. These
expressions simplify considerably when ${\gamma=0}$ to
\begin{eqnarray}
&& F_{q\,t}=e\,,\qquad  F_{ y\,p}=-g\,p^{-2}\,,\nonumber\\
&& \tilde{F}_{q\,t}=g\,,\qquad  \tilde{F}_{ y\,p}=+e\,p^{-2}\,.
\end{eqnarray}
The dual $\tilde{F}$ is obviously obtained from $F$ just by
interchanging ${e\rightarrow g}$, ${g\rightarrow-e}$, i.e.,
there is a duality between the \emph{electric charge $e$} and
the \emph{magnetic charge $g$}. Such electromagnetic fields
\emph{diverge at} ${p=0}$, i.e., at the singularity of the
gravitational field given by the $B$-metrics (\ref{Bmetric}),
(\ref{nonExpMetricBL}). In this case
\begin{equation}
\om = (e+\im\,g)\,\d \big(q\,\d t+\im\,p^{-1} \d y\big)
    = (e+\im\,g)\,\big(\d q\wedge\d t+\im\,p^{-2}\d y\wedge \d p \big)\,,
\end{equation}
and the corresponding 4-potential ${A=A_\mu\,\d x^\mu}$, such
that ${F=\d A}$, has a very simple form
\begin{equation}
A = e\,q\,\d t-g\,p^{-1}\d  y\,.
\end{equation}

Returning now to the most general case with $\gamma$, we can
 express the general electromagnetic field
(\ref{eq:PDnexF}) with respect to the null tetrad
(\ref{Tetradnonexp}) where now ${\alpha=0}$, ${\omega=1}$. In
the \emph{NP formalism} this is given by three complex
functions $\Phi_A$ defined as
\begin{equation}
\Phi_0=F_{\mu\nu}\,k^\mu\,m^\nu\,,\quad
\Phi_1={\textstyle\frac{1}{2}}F_{\mu\nu}\left(k^\mu\,l^\nu+\bar{m}^\mu\,m^\nu\right)\,,\quad
\Phi_2=F_{\mu\nu}\,\bar{m}^\mu\,l^\nu\,.
\end{equation}
They take the form
\begin{equation}
\Phi_1=-\frac{e+\im\,g}{2(\gamma+\im\,p)^2}\,,\qquad \Phi_0=0=\Phi_2\,,
\end{equation}
corresponding to the only non-vanishing tetrad components
\begin{equation}
F_{kl} \equiv F_{\mu\nu}\,k^\mu\,l^\nu =-\frac{e(\gamma^2-p^2)+2g\,\gamma\,p}{(\gamma^2+p^2)^2}\,,\quad
F_{\bar{m}m}\equiv F_{\mu\nu}\,\bar{m}^\mu\,m^\nu=-\im\,\frac{g(\gamma^2-p^2)-2e\,\gamma\,p}{(\gamma^2+p^2)^2}\,.\nonumber
\end{equation}
Since the related \emph{Ricci tensor} in the NP formalism is
${\Phi_{AB}=2\,\Phi_A\bar{\Phi}_B}$ the only non-vanishing
component is
\begin{equation}
\Phi_{11}= \frac{e^2+g^2}{2(p^2+\gamma^2)^2}\,,
\end{equation}
which is fully consistent with (\ref{nonExpMetric3NP}).

Finally, a \emph{complex invariant of the electromagnetic
field} reads
\begin{equation}
{\textstyle\frac{1}{8}}\,(F_{\mu\nu}\,F^{\mu\nu}+\im\,F_{\mu\nu}\,\tilde{F}^{\mu\nu})
\equiv \Phi_0\,\Phi_2-(\Phi_1)^2
=-\frac{1}{4}\frac{(e+\im\,g)^2}{(\gamma+\im\,p)^4}\,.
\label{invariantEM}
\end{equation}
It is non-zero, so that the \emph{electromagnetic field is
non-radiating} (non-null). Indeed, (since only ${\Phi_1\ne0}$)
it is of a \emph{general algebraic type} with the null vectors
$\boldk$ and $\boldl$ of the electromagnetic field aligned with
the double degenerate principal null directions of the Weyl
tensor representing type~D gravitational field.

Moreover, it can be seen from (\ref{invariantEM}) that for
${\gamma\neq0}$ the electromagnetic field is \emph{everywhere
finite}, and for ${p\to \infty}$ the field \emph{vanishes
asymptotically}. Only for the family of $B$-metrics (if, and
only if, ${\gamma=0}$) there is a singularity located at
${p=0}$.

\newpage

\section{Summary and conclusions}

We have here presented and analyzed the complete family of
non-expanding Pleba\'nski--Demia\'nski space-times which are
(electro)vacuum solutions with any cosmological constant of
algebraic type~D. Such a family can be explicitly obtained by
performing a specific limit (Section~\ref{intro}) leading to a
vanishing expansion, twist and shear, i.e., to the Kundt class
(Section~\ref{Kundt}). By demonstrating
(Section~\ref{removingalpha}) that the parameter $\alpha$,
originally representing acceleration, can always be removed, and
$\omega$ set to 1, we proved that all such solutions can be
written in the metric form (\ref{nonExpMetric3}),
(\ref{coeffnonexp3}). The only exception is the family of
direct-product geometries that is obtained by another limit
when ${\alpha=0=\omega}$ (Section~\ref{furthercases}).

This class of solutions contains two discrete parameters
${\epsilon_0,\epsilon_2=+1, 0, -1}$ and five continuous
parameters $n$, $\gamma$ and $e$, $g$, $\Lambda$. In our
contribution we thoroughly investigated the geometrical and
physical meanings of all these parameters, and we have provided
basic interpretations of the corresponding space-times.

First, in Section~\ref{Minkowski} we determined the geometrical
meaning of $\epsilon_0$ and $\epsilon_2$ in the case when all
other parameters are set to zero. We showed that these discrete
parameters correspond to specific new coordinate
representations of certain regions of the background Minkowski
space. In the presence of a cosmological constant
${\Lambda\not=0}$ the parameters $\epsilon_0$ and $\epsilon_2$
analogously determine specific coordinate representations of
the de~Sitter or anti-de~Sitter backgrounds, see
Section~\ref{deSitter} and our previous work \cite{PodHru17}.

The physical meaning of the parameter~$n$ was elucidated in
Section~\ref{B-metrics}. Its presence defines the family of
$B$-metrics, with a curvature singularity at ${p=0}$. In
particular, the $BI$-metric defined by ${\epsilon_2=1}$
represents an exact gravitational field of a tachyon of
``mass''~$n$, moving with an infinite velocity along a straight
line. The same physical interpretation can be given to the
$B$-metrics with $\Lambda$, in which case the tachyonic source
moves in the (anti-)de~Sitter universe, see
Section~\ref{B-metricsLambda}. On the other hand, the
$BIII$-metrics are special cases of the Levi-Civita and
Linet--Tian metrics for which ${\sigma=1/4}$, or its dual
${\sigma=0}$.

The meaning of the parameter  $\gamma$ was identified in
Section~\ref{interp_gamma} as a formal analogue of the NUT
parameter. Its presence in the most general vacuum metric
(\ref{nonExpMetric4}), (\ref{coeffnonexp4}) of the
non-expanding Pleba\'nski--Demia\'nski class (which contains
the parameters ${\epsilon_0,\epsilon_2, \Lambda, n,\gamma}$)
causes the curvature singularity of the generalized $B$-metrics
to be removed. As a by-product we also found two completely new
diagonal metric forms of the anti-de~Sitter space, namely
(\ref{adSgammaMetric1}) and (\ref{adSgammaMetric2}).

Finally, as shown in Section~\ref{interp_eg}, the additional
two parameters $e$ and $g$ denote electric and magnetic charge
parameters, respectively. The corresponding space-times
(\ref{nonExpMetric3F}), (\ref{coeffnonexp3F}) of ``charged
$B$-metrics with~$\Lambda$ and $\gamma$''  contain a specific
(source-free) electromagnetic field. We presented the explicit
form of this non-null Maxwell field (\ref{eq:PDnexF}),
(\ref{eq:PDnexFt}), and we described its properties.

We hope that this clarification of all the parameters of the
full family of non-expanding Pleba\'nski--Demia\'nski
(electro)vacuum solutions will help in finding useful
applications of this large and interesting class of exact
space-times.

\section*{Acknowledgements}

This work was supported by the Czech Science Foundation grant
GA\v{C}R 17-01625S. O.H. also acknowledges the support by the
Charles University Grant GAUK~196516. We are grateful to
Robert~\v{S}varc for reading the manuscript and some useful
suggestions.


\begin{thebibliography}{99}

\bibitem{PleDem76} Pleba\'nski, J. F. and Demia\'nski, M.
    (1976). Rotating charged and uniformly accelerating mass in
    general relativity, {\sl Ann. Phys.} {\bf 98}, 98--127.

\bibitem{GriPod06b} Griffiths, J. B. and Podolsk\'y, J. (2006).
    A new look at the Pleba\'nski--Demia\'nski family of
    solutions, {\sl Int. J. Mod. Phys. D} {\bf 15}, 335--369.

\bibitem{GriPod09} Griffiths, J. B. and Podolsk\'y, J. (2009).
    {\sl Exact Space-Times in Einstein's General Relativity}
    (Cambridge University Press, Cambridge).

\bibitem{GriPod05} Griffiths, J. B. and Podolsk\'y, J. (2005).
    Accelerating and rotating black holes, {\sl Class. Quantum
    Grav.} {\bf 22}, 3467--3479.

\bibitem{Carter68} Carter, B. (1968). Hamilton--Jacobi and
    Schr\"odinger separable solutions of Einstein's equations,
    {\sl Commun. Math. Phys.} {\bf 10}, 280--310.

\bibitem{Plebanski75} Pleba\'nski, J. F. (1975). A class of
    solutions of the Einstein--Maxwell equations, {\sl Ann.
    Phys.} {\bf 90}, 196--255.

\bibitem{Plebanski79} Pleba\'nski, J. F. (1979). The
    nondiverging and nontwisting type D electrovac solutions
    with~$\lambda$, {\sl J. Math. Phys.} {\bf 20}, 1946--1962.

\bibitem{GarPle82} Garc\'{\i}a D\'{\i}az, A. and Pleba\'nski,
    J. F. (1982). Solutions of type D possessing a group with
    null orbits as contractions of the seven-parameter
    solution, {\sl J. Math. Phys.} {\bf 23}, 1463--1465.

\bibitem{Kinnersley69} Kinnersley, W. (1969). Type D vacuum
    metrics, {\sl J. Math. Phys.} {\bf 10}, 1195--1203.

\bibitem{VandenBergh17} Van den Bergh, N. (2017).
    Algebraically special Einstein--Maxwell fields,
    {\sl  Gen. Relativ. Grav.} {\bf 49}, 9 (16pp).

\bibitem{PodHru17} Podolsk\'y, J. and Hru\v ska, O. (2017).
    Yet another family of diagonal metrics for de~Sitter and anti-de~Sitter spacetimes,
    {\sl Phys. Rev. D} {\bf 95}, 124052 (29pp).

\bibitem{EhlersKundt62} Ehlers, J. and Kundt, W. (1962).
    Exact solutions of the gravitational field equations, in
    {\sl Gravitation: An introduction to current research}
    (Wiley, New York), 49--101.

\bibitem{Martins96} Martins, M. A. P. (1996). The sources of
    the A and B degenerate static vacuum fields,
    {\sl  Gen. Relativ. Grav.} {\bf 28}, 1309--1320.

\bibitem{Gott74} Gott, J. R. (1974). Tachyon singularity: A
    spacelike counterpart of the Schwarzschild black hole,
    {\sl Nuovo Cimento B}, {\bf 22}, 49--69.

\bibitem{Linet} Linet, B. (1986). The static, cylindrically
    symmetric strings in general relativity with cosmological
    constant, {\sl J. Math. Phys.} {\bf 27}, 1817--1818.

\bibitem{Tian} Tian, Q. (1986). Cosmic strings with
    cosmological constant, {\sl Phys. Rev. D} {\bf 33}, 3549--3555.

\bibitem{GPLT} Griffiths, J. B. and Podolsk\'y, J.
    (2010). The Linet--Tian solution with a positive cosmological constant in four and
    higher dimensions, {\sl Phys. Rev. D} {\bf 81}, 064015 (6pp).

\bibitem{Hruska2015} Hru\v ska, O. (2015).
    The study of exact spacetimes with a cosmological constant.
    Diploma Thesis, Charles University, Faculty of Mathematics and Physics, Prague (193pp).

\end{thebibliography}
\end{document}